\documentstyle[12pt,times]{article}
\newlength{\abstractwidth}
\renewcommand{\thefootnote}{\fnsymbol{footnote}}
\renewcommand{\thanks}[1]{\footnote{#1}} % Use this for footnotes
\newcommand{\starttext}{
\setcounter{footnote}{0}
\renewcommand{\thefootnote}{\arabic{footnote}}}

\newcommand{\be}{\begin{equation}}
\newcommand{\bea}{\begin{eqnarray}}
\newcommand{\eea}{\end{eqnarray}}
\newcommand{\beq}{\begin{equation}}
\newcommand{\ee}{\end{equation}}
\newcommand{\eeq}{\end{equation}}

\newcommand{\<}{\langle}
\newcommand{\ophi}{{\cal O}_\phi}
\newcommand{\oc}{{\cal O}_C}
\renewcommand{\>}{\rangle}
\def\ba{\begin{eqnarray}}
\def\ea{\end{eqnarray}}
\newcommand{\PSbox}[3]{\mbox{\rule{0in}{#3}\includegraphics{#1}\hspace{#2}}}

\def\half{{1 \over 2}}
\def\d{{d \over 2}}

\def\AdS{{\rm AdS}}
\def\dmu{\partial _\mu}
\def\dnu{\partial _\nu}
\def\dmup{\partial _{\mu '}}
\def\dnup{\partial _{\nu '}}
\begin{document}
\begin{titlepage}
\bigskip
\hskip 3.7in\vbox{\baselineskip12pt
\hbox{MIT--CTP--2843}
\hbox{UCLA/99/TEP/2}
\hbox{hep-th/9903196}}
\bigskip\bigskip\bigskip\bigskip

\centerline{\Large \bf Graviton exchange and complete 4--point functions}
\smallskip
\centerline{\Large \bf in the AdS/CFT correspondence}

\bigskip\bigskip
\bigskip\bigskip

\centerline{ Eric D'Hoker$^{a}$, Daniel Z. Freedman$^{b,c}$, 
Samir D. Mathur$^{b}$, }
\medskip
\centerline{Alec Matusis$^{b}$  
and Leonardo Rastelli$^{b,}$\footnote[1]{\tt dhoker@physics.ucla.edu, dzf@math.mit.edu, me@ctpdown.mit.edu, alec\_m@ctp.mit.edu, rastelli@ctp.mit.edu.}}
\bigskip
\bigskip
\centerline{$^a$ \it Department of Physics}
\centerline{ \it University of California, Los Angeles, CA 90095}
\bigskip
\centerline{$^b$ \it Center for Theoretical Physics}
\centerline{ \it Massachusetts Institute of Technology}
\centerline{ \it Cambridge, {\rm MA}  02139}
\bigskip
\centerline{$^c$ \it Department of Mathematics}
\centerline{ \it Massachusetts Institute of Technology}
\centerline{\it Cambridge, {\rm MA} 02139}
\bigskip\bigskip

\begin{abstract}
The graviton exchange diagram for the correlation function
of  arbitrary scalar operators is evaluated in anti--de Sitter
space, $AdS_{d+1}$. This enables us to complete the computation
of the 4--point amplitudes of dilaton and axion fields 
in IIB supergravity
on $AdS_5 \times S_5$.   
By the AdS/CFT correspondence, we obtain the 4--point functions
of the marginal  operators ${\rm Tr }( F^2+\dots)$ and ${ \rm Tr}( F \tilde{F} +\dots)$
 in ${\cal N}=4$, $d=4$ $SU(N)$ SYM at large $N$, large $g_{YM}^2 N$.
The short distance asymptotics of the amplitudes are studied.
We find that in the direct channel the leading power singularity
agrees with the expected contribution of the stress--energy
tensor in a double OPE expansion. Logarithmic singularities
occur in the complete 4--point functions at subleading orders.

\end{abstract}

\end{titlepage} 
\starttext
\baselineskip=18pt
\setcounter{footnote}{0}
\section{Introduction }

The calculation of correlation functions is one 
useful way to test and explore the AdS/CFT
correspondence \cite{maldacena,polyakov,witten}, which relates
$d$--dimensional conformal field theories with compactifications of
string/M theory involving $AdS_{d+1}$. The simplest example of the
correspondence is the duality between ${\cal N}=4,\; d=4\; SU(N)$ SYM
theory and type IIB string theory on $AdS_5\times S^5$ with $N$ units
of 5--form flux and compactification radius $R^2=\alpha'\: (g_{YM}^2
N)^{1\over 2}$. In the large $N$ limit with $\lambda=g_{YM}^2
N$ fixed and large the supergravity approximation is
valid. Correlators of gauge invariant local operators in the CFT at
large $N$ and strong t'Hooft coupling $\lambda$ are related to
supergravity amplitudes according to the prescription of
\cite{polyakov,witten}. The 5--dimensional Newton constant $G_5\sim
{R^3/ N^2}$, so that the perturbative expansion in supergravity,
if ultraviolet convergent, corresponds to the $1/N$ expansion
in the CFT.

Broadly speaking, 2-- and 3--point functions 
(see {\it e.g.} \cite{march,seiberg, dhfskiba})
have provided evidence that
the conjectured correspondence is correct, but 4--point functions are
expected to contain more information about the non-perturbative dynamics
of the CFT. Previous studies relevant to 4--point correlators include \cite{canadians}--\cite{other4points}. 
4--point
correlators for 
contact interactions of
scalars in the bulk theory were the first to be studied
\cite{canadians,
 liutseytlin2,
august} followed by diagrams with exchanged gauge bosons \cite{dhfgauge} 
and scalars \cite{talk, liu, dhfscalar}. (See also \cite{chsch1,chsch2} for
a 
different approach). $\alpha'/R^2$ corrections are considered in
\cite{brodie}, and there is an extensive literature 
on instanton contributions, see {\it e.g.} \cite{instantons}.

The simplest 4--point correlators that can be studied are
those involving the marginal operators ${\cal O}_\phi \sim {\rm Tr}
(F^2+\dots)$ and $\oc \sim {\rm Tr}
(F\tilde F+\dots) $ corresponding to the dilaton and axion
supergravity fields, as first stressed in \cite{liutseytlin2}. To
leading order in $N$, the amplitudes $\< \ophi\ophi\ophi\ophi\>$, $\< \oc\oc\oc\oc\>$ and $\< \ophi\oc\ophi\oc\>$
factorize in products of 2--point
functions (see Figures 1a and 3). Thanks to the non--renormalization theorem
for the 2--point functions \cite{anselmi,march}, these disconnected
contributions do not 
receive corrections in powers of $\alpha'/R^2 = 1/\lambda^{1/2}$.
 The next contribution to the 4--point amplitudes 
is thus a $1/N^2$ effect
and involves tree--level, connected supergravity diagrams like the
ones in Figure 2. The computation of $\< \ophi\ophi\ophi\ophi\>$,
$\< \oc\oc\oc\oc\>$ and
$\< \ophi\oc\ophi\oc\>$ was started in \cite{august} with the
evaluation of the relevant quartic and scalar exchange diagrams
(Figure 2s,u,q and Figure 4). Here we complete the computation by 
evaluating the
remaining graviton exchange diagram (Figure 2t) and we initiate the
analysis of the first realistic 4--point amplitude in the AdS/CFT
correspondence.

We also present what we believe is a cross--checked and
reliable  calculation of the graviton
exchange diagram between pairs of external scalars of arbitrary mass
in $AdS_{d+1}$ for arbitrary $d$. The calculation 
was facilitated by the recently derived covariant form of the
graviton propagator \cite{january}, but it is still very 
complex compared to earlier
work.

One theoretical framework to analyze results on 4--point functions
in the operator product expansion (OPE) \cite{talk, ferraraope}. 
The mere assumption of an OPE
is  quite restrictive and imposes constraints on
the allowed form of the result. Let us assume a double  ``t--channel'' OPE
of the schematic form
\be \label{ope}
 \< {\cal O}_1(x_1) {\cal O}_2(x_2) {\cal O}_3(x_3) {\cal O}_4(x_4) 
\> =\\ 
 \sum_{n,m}\,\frac{\alpha_{n}
\;\< {\cal O}_n(x_1){\cal O}_m(x_2) \>
\;\beta_m }{(x_1-x_3)^{\Delta_1+\Delta_3-\Delta_m}(x_2-x_4)^{\Delta_2+\Delta_4-\Delta_n}} 
\ee
containing the contribution of various primary operators ${\cal O}_p$
and 
their descendents $\bigtriangledown^k {\cal O}_p$ in the intermediate state. 
For simplicity we have assumed that these are scalars, but vector and
tensor operators contribute in a similar way, each with a
characteristic tensor structure. (For primary operators,
$\< {\cal O}_p {\cal O}_{p'} \>$ vanishes unless $\Delta_p = \Delta_{p'}$).

Recognizing in the supergravity 4--point results a structure of the
form (\ref{ope}) should allow to determine the operator content of the
theory and its OPE structure in the large $N$, large $\lambda$ limit.
Preliminary computations \cite{talk,liu} have indicated that
the supergravity diagrams contain the expected contributions to
(\ref{ope}) of chiral primary operators and their superconformal
descendents. It is however clear that these contributions alone do
not reproduce the supergravity result \cite{august}.
A natural expectation is that appropriately defined 
normal--ordered {\it products} of chiral primaries and
descendents also contribute to the OPE 
and form the full operator content of the theory in this
limit. This set of operators has a dual interpretation in terms of
multi--particle Kaluza--Klein states in supergravity. Massive string
states are expected to decouple in this limit\footnote{Group--theoretic aspects
of multi--particle and string states have been considered in \cite{group}.}.
The computation of a
complete realistic 4--point correlator presented here should allow to
put  these ideas to test.

An interesting issue raised in \cite{august} is the presence in the
4--point supergravity amplitudes of logarithms of the coordinate
separation between two points in the limit when the points come
close. Logarithmic singularities appear to be a generic feature of all
the AdS processes studied so far \cite{dhfgauge,liu,dhfscalar}, and
we find the same situation for the graviton exchange. The question
then is whether the logarithms cancel
when the various contributions to a realistic correlator are assembled. 
If not, we should ask whether the
logarithms can still be incorporated in the OPE framework. 
Here we find that logarithmic singularities do indeed occur in the
complete 4--point functions.

As pointed
out by Witten \cite{wittenpc}, logarithms can generically arise in the
perturbative expansion of a CFT 4--point correlator 
as renormalization effects like 
mixings and corrections to the dimensions of the exchanged
operators. The perturbative parameter is in this case $1/N$,
which is mapped by the correspondence to
the gravitational coupling constant.
The operators $\ophi$ and $\oc$ are chiral and hence their dimensions
are protected, but their OPE's contain (besides chiral
contributions like the stress--energy tensor) non--chiral
composite operators like 
$:\ophi \ophi:$ that require  a careful definition and can lead 
to renormalization effects \cite{wittenpc}. (A somewhat different viewpoint
has been described in a very recent
paper \cite{kogan}, see also \cite{lcft}).  

It is an interesting
subject for future work to analyze the constraints imposed by this
interpretation on the allowed form of the logarithmic
singularities 
and to assess 
the compatibility of these constraints with the supergravity results.

The paper is organized as follows. 

In Section 2,
we present the supergravity graphs that contribute
to 4--point functions involving $\ophi$, $\oc$,  summarize
our results for the amplitudes and make some remarks about their
OPE interpretation.

In Section 3,
we describe the general set--up for the calculation of
the graviton exchange amplitude. We give a few geometric
identities, summarize the results for the scalar and
graviton propagators and present
the integral associated with the graviton exchange graph.

In Section 4 and Section 5 we separately describe two independent
computations of the graviton amplitude,
for $\Delta = \Delta' =d=4$ in Section 4 and for general
$\Delta$, $\Delta'$ and $d$ in Section 5. Both computations
reduce
the graviton exchange amplitude to  finite
sums of scalar quartic graphs (see Figure 6). The two results
are shown to precisely agree for $\Delta = \Delta'=d=4$.

In Section 6, we develop integral representations and asymptotic
series
expansions for the quartic graphs (Figure 5), which are
the basic building blocks of the answer.
We find asymptotic serieses for the graviton
exchange in terms of two conformally invariant 
variables.

Finally, in the Appendix 
we discuss  some properties and mathematical
identities of the quartic graphs.

\newpage
\setcounter{equation}{0}

\section{4--point functions in the dilaton--axion sector}

Following \cite{liutseytlin2}, we first discuss  the dilaton--axion--graviton
sector of IIB supergravity, dimensionally reduced on the
classical background solution $AdS_5 \times S_5$ 
keeping only the
constant modes on $S_5$. The relevant part of  5--dimensional
action is\footnote{The metric appearing in (\ref{action}) is
{\it not}  the restriction  of the original 10--dimensional
metric to $AdS_5$, but it is related to it by a Weyl rescaling of the 
metric fluctuations~\cite{vannieuv, liutseytlin2}.
The fluctuation $h'_{\mu \nu}$ that gives the massless graviton
in $AdS_5$ is given in terms of  the original $h_{\mu \nu}$ by $h_{\mu \nu}=
h'_{\mu \nu} -\frac{1}{3} \bar g_{\mu \nu} h^\alpha_{\, \alpha}$,
where $\alpha$ is an index along $S_5$ and $\bar g_{\mu \nu}$
the background metric \cite{vannieuv}. 
%For notational convenience,
%we will henceforth drop primes, $h'_{\mu \nu} \rightarrow h_{\mu \nu}$.
} 
\be \label{action}
S = \frac{1}{2 \kappa^2} \int_{AdS_5} d^5z \,\sqrt{g} \;\left(
-{\cal R} +\frac{1}{2}\,g^{\mu \nu} \partial_\mu \phi \partial_\nu \phi
+\frac{1}{2}\,e^{2 \phi} \,g^{\mu \nu} \partial_\mu C \partial_\nu C
\right).
\ee
Th 5--dimensional gravitational coupling $\kappa$ is related
to the parameters of the compactification by 
$2 \kappa^2 = \frac{15 \pi^3 R^3}{N^2}$, where $N$ is the number of units
of  5--form flux and  $R$ the radius of the 5--sphere (equal 
to the $AdS_5$ scale, see 
equ.(\ref{adsmetric}) below).  We will usually set
the $AdS_5$ scale $R \equiv 1$.

\subsection{Witten diagrams}

%%%%%%%%%%%%%%%%%%%%%%%%%%%%%%%%%%figure%%%%%%%%%%%%%%%%%%%%%%%%%%%%%%%%%%%

\begin{figure} 
\begin{center} \PSbox{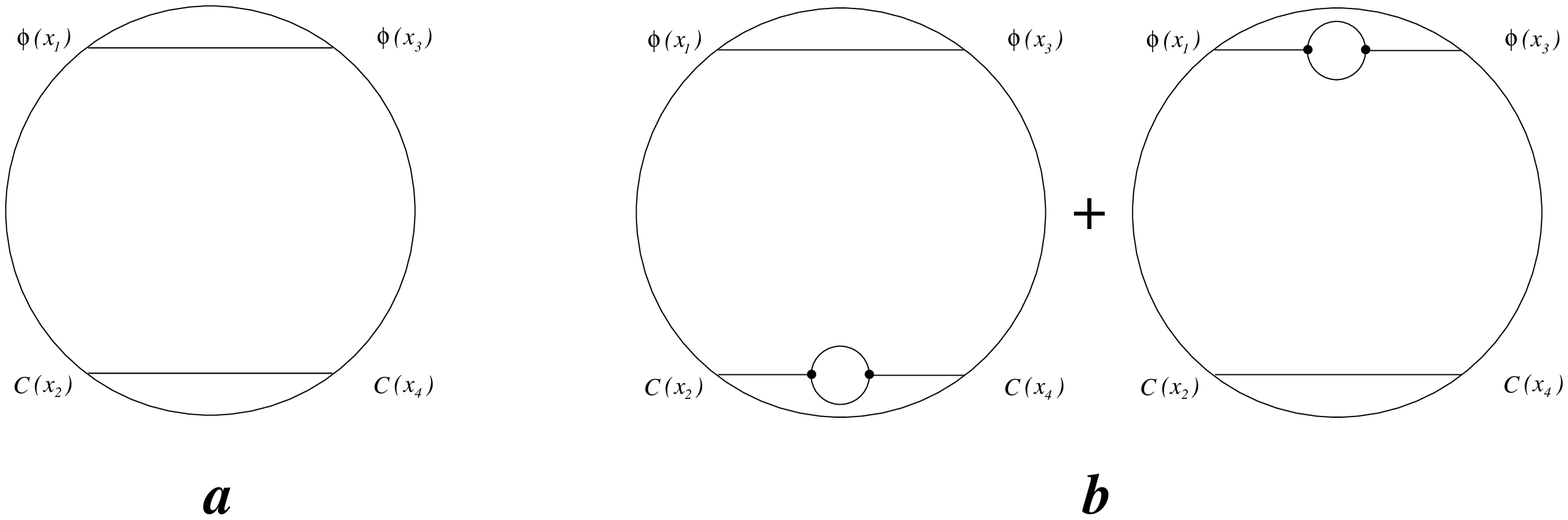 hscale=55 vscale=55}{5.1in}{1.3in} \end{center}
\caption{Disconnected contribution  to $\<\ophi \oc \ophi \oc \>$. $a$: $O(N^4)$;   $b$:
 $O(N^2)$.}
\end{figure}

%%%%%%%%%%%%%%%%%%%%%%%%%%%%%%endfigure%%%%%%%%%%%%%%%%%%%%%%%%%%%%%%%%%%%

%%%%%%%%%%%%%%%%%%%%%%%%%%%%%%%%%%figure%%%%%%%%%%%%%%%%%%%%%%%%%%%%%%%%%%%

\begin{figure} 
\begin{center} \PSbox{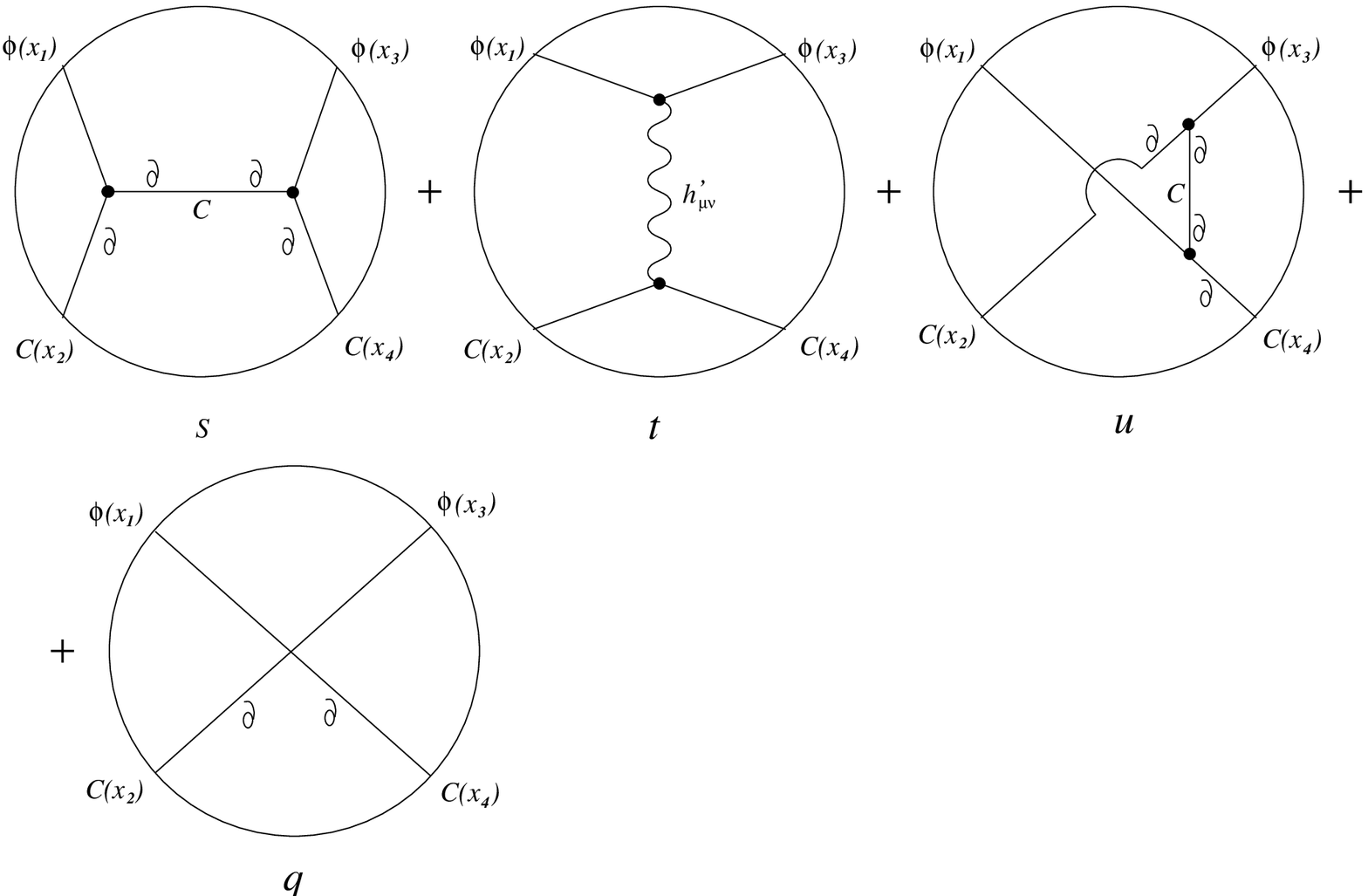 hscale=70 vscale=70}{4.6in}{3.3in} \end{center}
\caption{Connected $O(N^2)$ contributions to  $\<\ophi \oc \ophi \oc \>$.}
\end{figure}

%%%%%%%%%%%%%%%%%%%%%%%%%%%%%%endfigure%%%%%%%%%%%%%%%%%%%%%%%%%%%%%%%%%%%
We wish to implement the prescription of \cite{polyakov,witten}
to compute the CFT correlators $\< \ophi \oc \ophi \oc \>$,
 $\< \ophi \ophi \ophi \ophi \>$, $\< \oc \oc \oc \oc \>$, where
$\ophi \sim  {\rm Tr} (F^2 +...)$,
$\oc \sim {\rm Tr} (F \tilde F +...)$ are the exactly marginal ($\Delta =4$)
SYM 
operators corresponding to the 
dilaton and axion fields~\cite{witten}\footnote{The precise structure of the 
composite operators 
$\ophi$ and $\oc$ in terms
of elementary SYM fields is in principle given by the 
variation of the on--shell
${\cal N}=4$ lagrangian with respect to the marginal couplings
$g_{YM}$ and $\theta$, or by supersymmetry transformations
starting from the chiral primary ${\rm Tr} X^{(i} X^{j)}$.}.

Let us first consider $\< \ophi(x_1) \oc(x_2) \ophi(x_3) \oc(x_4) \>$.
The leading large $N$ contribution is given by the disconnected
diargam in Figure 1a. This diagram, being the product of two 2--point
functions,  is proportional to $N^4/(x_{13}^8x_{24}^8)$. 

The next
contribution, of order $N^2$, comes from the diagrams in Figures 1b and
2. However, the one--loop diagrams in Figure 1b, thanks to the fact
that the dimensions of the chiral operators $\ophi,\oc$ are protected,
only give a $1/N^2$ correction to the overall coefficient of the
amplitude in Figure 1a\footnote{This correction precisely accounts for
the fact the gauge group is $SU(N)$ rather than $U(N)$. Note that 
validity of the correspondence seems to require that there are no 
higher loop corrections in the supergravity 2--point functions.}. Among the diagrams in Figure 2, the sum s+u+q
has been computed in \cite{august}. 

Sections 4 and 5 of the paper are
devoted to evaluation of the remaining graviton exchange diagram t.

Similarly, Figures 3 and 4 reproduce the relevant diagrams for 
$\< \oc(x_1) \oc(x_2) \oc(x_3) \oc(x_4) \>$. The connected diagrams
for $\< \ophi(x_1) \ophi(x_2) \ophi(x_3) \ophi(x_4) \>$ involve only
graviton exchanges. As shown in \cite{august} the s,t,u scalar exchange
diagrams in Figure 4 add up to zero. Hence, to this order, 
\be
\< \ophi \ophi \ophi \ophi \>= 
\< \oc \oc \oc \oc \>.
\ee
\newpage
%%%%%%%%%%%%%%%%%%%%%%%%%%%%%%%%%%figure%%%%%%%%%%%%%%%%%%%%%%%%%%%%%%%%%%%

\begin{figure} 
\begin{center} \PSbox{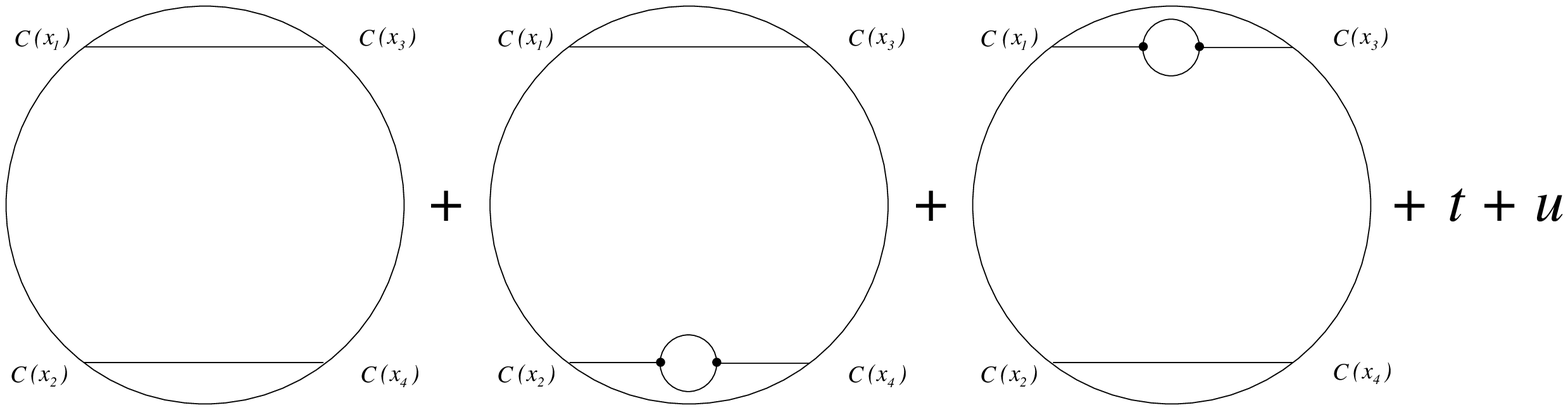 hscale=55 vscale=55}{4.6in}{1.3in} \end{center}
\caption{Disconnected $O(N^4)$ and $O(N^2)$ contributions to $\<\oc \oc \oc \oc \>$.}
\end{figure}

%%%%%%%%%%%%%%%%%%%%%%%%%%%%%%endfigure%%%%%%%%%%%%%%%%%%%%%%%%%%%%%%%%%%%

%%%%%%%%%%%%%%%%%%%%%%%%%%%%%%%%%%figure%%%%%%%%%%%%%%%%%%%%%%%%%%%%%%%%%%%

\begin{figure} 
\begin{center} \PSbox{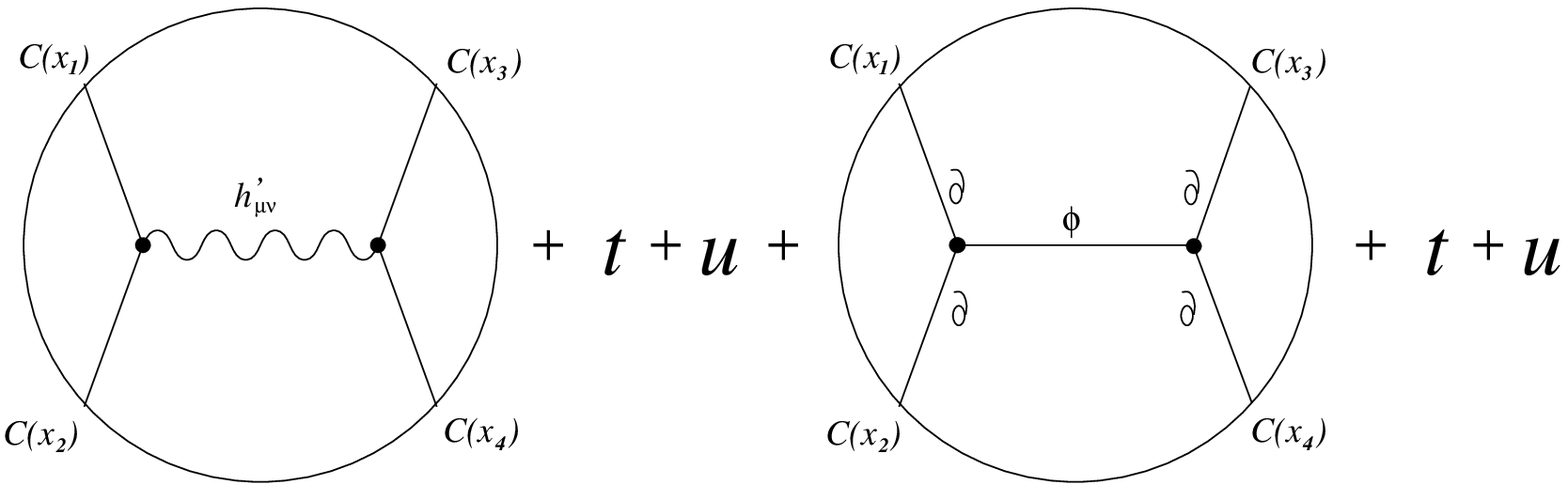 hscale=70 vscale=70}{4.6in}{1.3in} \end{center}
\caption{Connected $O(N^2)$ contributions to  $\<\oc \oc \oc \oc \>$.}
\end{figure}

%%%%%%%%%%%%%%%%%%%%%%%%%%%%%%endfigure%%%%%%%%%%%%%%%%%%%%%%%%%%%%%%%%%%%

\subsection{Summary of results}
It turns out that upon integration over one of the bulk points, all
4--point AdS processes with external scalars, including the graviton 
exchange, reduce to a finite sum of
scalar quartic graphs (see Figure 6). We denote
quartic graphs of external conformal dimensions $\Delta_i$ with
the symbol
 $D_{\Delta_1\Delta_3\Delta_2\Delta_4}(x_1,x_3,x_2,x_4)$,  
as in Figure 5 (see equation (\ref{AD}) for the precise definition and 
the Appendix for a discussion of properties
of these functions). 

The final result for the graviton exchange graph in Figure 2t as sum
of quartic graphs (for
$\Delta=\Delta'=d=4$), derived in Sections 4 and 5 below, is
\bea \label{gravresult}
I_{\rm grav}& =&
\left( \frac{6}{\pi^2} \right)^4
\left[ 16\, x_{24}^2\left({1\over 2s}-1\right)\,
D_{4455}+{64\over 9}{x_{24}^2\over x_{13}^2}{1\over s}\, D_{3355}+
{16\over 3}{x_{24}^2\over x_{13}^4}{1\over s}\, D_{2255}
\right. \\
&&\left.+18\,D_{4444}-
{46\over 9\,x_{13}^2}\,D_{3344}-{40\over 9\,x_{13}^4}\,D_{2244} -{8\over 3\,x_{13}^6}\,D_{1144} \right]
 \,,
\nonumber\eea
where we have introduced the conformally invariant variable 
\be 
s \equiv
\frac{1}{2} \frac{x_{13}^2 x_{24}^2}{x_{12}^2 x_{34}^2 + x_{14}^2
x_{23}^2}.
\ee
See equations (\ref{(3.22)}, \ref{(4.18result)}--\ref{(4.19)}) for the analogous result in the general case of
arbitrary $\Delta,\Delta',d.$

We also recall the result \cite{august} for the sum of the amplitudes 
s, q, u in Figure 2
\be\label{suqresult}
I_{s}+I_{u}+I_{q}=\left( \frac{6}{\pi^2} \right)^4\left[64\,
x_{24}^2D_{4455}-32\,D_{4444}\right]\,.
\ee

The sum of (\ref{gravresult}) and (\ref{suqresult}) gives the
connected order $N^2$ contribution to the correlator
$\< \ophi(x_1) \oc(x_2) \ophi(x_3) \oc(x_4) \>$. 
The analogous result
for $\< \ophi \ophi \ophi \ophi \>= 
\< \oc \oc \oc \oc \>$ is obtained by cross--symmetrization of
(\ref{gravresult}).
%%%%%%%%%%%%%%%%%%%%%%%%%%%%%%%%%%figure%%%%%%%%%%%%%%%%%%%%%%%%%%%%%%%%%%%

\begin{figure} 
\begin{center} \PSbox{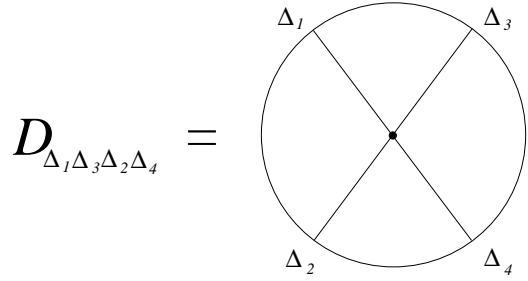 hscale=55 vscale=55}{2.7in}{1.3in} \end{center}
\caption{Definition of $D_{\Delta_1 \Delta_3 \Delta_2 \Delta_4}$.}
\end{figure}

%%%%%%%%%%%%%%%%%%%%%%%%%%%%%%endfigure%%%%%%%%%%%%%%%%%%%%%%%%%%%%%%%%%%%
The functions $D_{\Delta_1\Delta_3\Delta_2\Delta_4}$ admit simple
integral representations (see  Section 6.1) and can all be obtained
as derivatives with respect to $x_{ij}^2$ of a single function
(see Section A.3).
In Section 6 we develop asymptotic series expansions for
$D_{\Delta_1\Delta_3\Delta_2\Delta_4}$ in the  conformally
invariant variables $s$ and $t$,
\be
 t= \frac{x_{12}^2 x_{34}^2 -x_{14}^2 x_{23}^2}{x_{12}^2 x_{34}^2
+x_{14}^2 x_{23}^2} \,. 
\ee
We consider the ``direct'' or t--channel limit  $|x_{13}|\ll|x_{12}|,\;|x_{24}|\ll|x_{12}|$ which corresponds to $s,t\to 0$.
The singular power terms in this limit are given by
\be
I_{{\rm grav}}\bigg |_{{\rm sing}} = \label{Igravsing}
 {2^{10} \over  35 \pi ^6 } {1 \over x_{13}^8 x_{24}^8} \biggl [
s \,\bigl (7t^2 +6 t^4 \bigr )
+ s^2 \,\bigl (- 7-+3  t^2 \bigr )
- 8\,s^3  \biggr].
\ee
In addition, as in \cite{august,dhfgauge,liu,dhfscalar} we find an infinite series of terms logarithmic in $s$: 
\bea \label{Igravlogsummary}
I_{\rm grav} \bigg |_{\rm log}  & =& {3\cdot 2^{3} \over \pi ^6} {\ln s \over 
x_{13}^8x_{24}^8}
\sum _{k=0} ^\infty s^{4+k} \ {\Gamma (k+4) \over \Gamma (k+1)} \bigg \{
-2(5k^2 +20k+16)(3k^2+15k+22) a_{k+3}(t) \nonumber\\&&
 + (k+4)^2 (15k^2 +55 k^2 +42) a_{k+4}(t) \bigg \},
\eea
where the  functions $a_k(t)$ are given by
\be
a_k(t)=\int_{-1}^1\! d\lambda {(1-\lambda ^2) ^k \over (1+\lambda t)^{k+1}}=
\sqrt\pi{\Gamma(k+1) \over \Gamma\left(k+{3\over 2}\right)}\,
F\left({k+1\over 2},{k\over 2}+1;k+{3\over 2};t^2\right).
\ee 
As clear from the hypergeometric representation, $a_k(t)$ admit power series expansions in $t^2$ with radius of convergence 1.
Here we do not display the non--singular power terms 
in $I_{{\rm grav}}$  (see Section 6.2).

The analogous result for the sum of the graphs s+u+q in Figure 2 is
\be \label{Isuqlog}
I_{\rm s}+I_{\rm u}+I_{\rm q} \bigg |_{\rm log}={2^6 \cdot 3 \cdot 5 \over \pi^6}{\ln s\over x_{13}^8
x_{24}^8}\sum_{k=0}^\infty s^{k+4}\left\{(k+1)^2(k+2)^2(k+3)^2(3k+4)\:a_{k+3}(t)\right\}\,.
\ee
The contribution $I_{\rm s}+I_{\rm u}+I_{\rm q}$
has no power singularities.

We now turn to discuss some  physical implications of these results.

\subsection{OPE interpretation}

Let us compare the singular power terms of (\ref{Igravsing})
with those expected form the OPE (\ref{ope}). In the direct channel
limit $|x_{13}| \ll |x_{12}|$, $|x_{24}| \ll |x_{12}|$ the leading terms
of the variables $s$ and $t$ are
\be \label{stleading}
s \sim \frac{x_{13}^2 x_{24}^2}{4 \,x_{12}^4} \qquad t \sim - \frac{x_{13} \cdot J(x_{12}) \cdot x_{24}}{x_{12}^2}
\ee
where $J_{ij}= \delta_{ij} -2 y_i y_j /y^2$ is the well--known
Jacobian tensor of the conformal inversion $y'_i= y_i/y^2$. The leading
term of  (\ref{Igravsing}) can then be written as
\be \label{Igravsingleading}
I_{{\rm grav}}\bigg |_{{\rm sing}}= \frac{2^6}{5 \pi^6} \frac{1}{x_{13}^6 x_{24}^6}
\frac{4 (x_{13}\cdot J(x_{12}) \cdot x_{24} )^2 - x_{13}^2 x_{24}^2}{x_{12}^8} + \dots
\ee
with subleading terms suppressed by powers of $|x_{13}|/|x_{12}|$
and $|x_{24}|/|x_{12}|$. We note from (\ref{ope}) that
(\ref{Igravsingleading}) describes
the contribution to the OPE of an operator ${\cal O}_p$ of dimension
$\Delta=4$. We show below that the tensorial structure agrees with the
the expected contribuion of the stress--energy tensor of the boundary theory.
It is worth mentioning first that various subcontributions to the amplitude
$I_{{\rm grav}}$ (some of the $D$ functions in (\ref{gravresult})) have 
leading power $1/(x_{13}^6 x_{24}^6 x_{12}^4)$
indicative of a scalar operator of dimension $\Delta=2$, which would not
be expected in the graviton exchange process.  The fact that this term cancels in the full
amplitude is then an important check of the calculation.

Let us consider a scalar operator ${\cal O}_\Delta$ of scale--dimension
$\Delta$ in $d$--dimensional space--time. The contribution of the conserved
traceless stress--tensor $T_{ij}$ to the OPE of  ${\cal O}_\Delta(x_1) {\cal O}_\Delta(x_3)$
is
\be \label{OOT}
{\cal O}_\Delta(x_1) {\cal O}_\Delta(x_3) \sim k \; \frac{ x_{13 \,i}\, x_{13 \,j} }{\; x_{13}^{ 2 \Delta +2 -d } }
\; T_{ij}(x_1)
\ee
and the 2--point function of the stress tensor is
\be
\<T_{ij}(x_1) T_{kl}(x_2) \> = \frac{c}{2} \,
\frac{J_{ik}(x_{12}) J_{jl}(x_{12})+ J_{il}(x_{12}) J_{jk}(x_{12}) -\frac{2}{d} \delta_{ij} \delta_{kl}}{x_{12}^{2d}} 
\ee
which is conserved and traceless in any dimension.
Note that $J_{ik}(y) J_{kj}(y) = \delta_{ij}$. We thus see that the stress tensor
contribution to the general scalar double OPE is
\be
\< {\cal O}_{\Delta}(x_1) {\cal O}_{\Delta'}(x_2) {\cal O}_{\Delta}(x_3) {\cal O}_{\Delta'}(x_4)
\sim \frac{k c k'}{d} \,
\frac{d(x_{13} \cdot J(x_{12}) \cdot x_{24})^2 -x_{13}^2 x_{24}^2}{ {x_{13}^{2 \Delta +2 -d}}\,
\, {x_{24}^{2 \Delta' +2-d}} \, x_{12}^8 }
\,.
\ee 
This form is in perfect agreement with (\ref{Igravsingleading}).
Further relevant information on 2-- and 3--point functions of the stress--energy tensor
can be found in \cite{stress}.

Let us now consider the logarithmic terms. We see from the sum of (\ref{Igravlogsummary}) and (\ref{Isuqlog})
that an infinite series of terms logarithmic in $s$ occurs in the direct
channel expansion of $\< \ophi \oc \ophi \oc \>$. Since the serieses
 (\ref{Igravlogsummary}) and (\ref{Isuqlog}) have a rather different structure, this conclusion
appears quite robust. (In particular, it is insensitive to the relative normalization
of $I_{{\rm grav}}$ and $I_{{\rm s}}+I_{{\rm u}}+I_{{\rm q}}$). We plead exhaustion and excuse
ourselves from carrying a similar analysis for the crossed channel limit of 
$\< \ophi \oc \ophi \oc \>$ and for $\< \ophi \ophi \ophi \ophi \>$. The reader can find the necessary
ingredients in Section 6.2.
As mentioned in the Introduction, one should be able to interpret
these logarithmic terms as $1/N^2$ renormalization effects related
to the contribution of composite operators to the OPE (\ref{ope}) \cite{wittenpc}.
For example, the leading logarithmic term in the direct channel limit, $\frac{1}{(x_{12})^{16}} \log(\frac{x_{13} x_{24}}{x_{12}^2})$,
could be related to the presence
in (\ref{ope}) of the non--chiral composite operators $:\ophi \ophi:$ and
$:\oc \oc:$. It is an interesting  topic for
future research to precisely identify the contributions of various composite operators,
and the patterns of their renormalization and mixing, in the
intricate series structures (\ref{Igravlogsummary}), (\ref{Isuqlog}). A detailed
OPE intepretation of these supergravity results should provide us with 
 new non--perturbative information about the ${\cal N} =4$ SYM theory.

\section{General set--up}

As in most previous work on correlation functions, we work on the Euclidean
continuation of $\AdS_{d+1}$, viewed as the upper half
space in $z_\mu \in {\bf R}^{d+1}$, with $z_0 >0$. The metric $g_{\mu \nu}$
and Christoffel symbols $\Gamma ^\kappa _{\mu \nu}$ are given by 
\bea  \label{adsmetric}
  ds^2 & =& \sum _{\mu, \nu=0} ^ d g_{\mu \nu} dz_\mu dz_\nu
       ={R^2 \over z_0^2} (dz^2_0 + \sum ^d_{i=1} \, dz^2_i) 
\\
\Gamma ^\kappa _{\mu \nu} & 
 =& {1 \over{R\, z_0}} \bigl ( 
 \delta _0 ^\kappa \delta_{\mu \nu}- \delta _{\mu 0} \delta ^\kappa _\nu 
                   - \delta _{\nu 0} \delta ^\kappa _\mu \bigr )
\label{2.1}
\eea
and the curvature scalar is ${\cal R}=-d(d+1)/R^2$. We henceforth
set the AdS scale $R \equiv 1$.
This space is a maximally symmetric 
solution
of the gravitational action
\be
S_g = -{ 1 \over 2 \kappa ^2} \int dz \sqrt g ({\cal R} - \Lambda) 
\label{2.2}
\ee
with $\Lambda=-d(d-1)$.

It is well known that invariant bi--scalar functions on $\AdS_{d+1}$, 
such as scalar field
propagators, are most  simply expressed in terms of the  chordal distance 
variable $u$, defined by
\be
  u \equiv 
  {(z-w)^2 \over 2z_0 w_0}
\label{2.6}
\ee
where $(z-w)^2 = \delta_{\mu \nu}(z-w)_{\mu} (z-w)_{\nu}$ is the ``flat
Euclidean distance". Invariant tensor functions, such as the gauge or the
graviton propagator, may be expanded in terms of bases of invariant bi-tensors,
which are derivatives of  $u$. For example, for rank 1, we
have ($\dmu =\partial /\partial z^\mu$ and $\dnup = \partial /\partial
w^{\nu'}$)
\bea
  \partial_{\mu} u &= &{1 \over z_0}
     \left(\frac{(z-w)_{\mu}}{ w_0} - u \delta_{\mu 0}\right) \\
   \partial_{\nu'} u &=& {1 \over w_0}
    \left(\frac{(w-z)_{\nu'} }{z_0} - u \delta_{\nu' 0}\right) \, .
\label{2.7}
\eea
and for rank 2, there is $\partial_{\mu}u \partial_{\nu'}u$
as well as
\be
  \partial_{\mu} \partial_{\nu'}u= -
  {1 \over z_0w_0} \big [ \delta_{\mu \nu'} +
  {1 \over w_0} (z-w)_{\mu} \delta_{\nu'0} +
  {1 \over z_0} (w-z)_{\nu'}\delta_{\mu 0} -
  u \delta_{\mu 0} \delta_{\nu'0} ] \,.
\label{2.8}
\ee
Throughout this paper, we shall
also make use of differentiation and contraction relations between these
basis tensors, which we list here, 
\bea
     \Box u= D^{\mu} \partial_{\mu} u &=& (d+1)(1+u) 
\label{2.9a} \\
      D^{\mu} u \ \partial_{\mu} u &=& u (2+u) 
\label{2.9b} \\
      D_{\mu} \partial_{\nu} u &=& g_{\mu \nu}(1+u) 
 \label{2.9c} \\    
      (D^{\mu}u)\ (D_{\mu} \partial_{\nu}\partial_{\nu'}u)
          &= &\partial_{\nu}u \partial_{\nu'}u 
\label{2.9d} \\
      (D^\mu u) \ (\dmu \dnup u) & =& (1+u) \dnup u
 \label{2.9e} \\
      (D^\mu \dmup u) \ (\dmu \dnup u) & =& g_{\mu ' \nu '} + \dmup u \dnup u\, .
 \label{2.9f} \\
\Box F(u) & = & u (u+2) F''(u) + (d+1)(1+u) F'(u) \label{box}
\eea
These relations may be derived using (\ref{2.7}), (\ref{2.8}) and the metric 
and Christoffel symbols of (\ref{2.1}) for $\AdS_{d+1}$.

\subsection{Scalar and graviton propagators}

The bulk--to--boundary propagator (or Poisson kernel) for a scalar field
of mass $m^2 = \Delta (\Delta-d)$ is well--known \cite{witten, march}
 and given by
\be
K_\Delta (z, \vec{x}) = C_\Delta \tilde K_\Delta (z, \vec x) =
C_\Delta \left ( { z_0 \over z_0 ^2 +
(\vec{z}-\vec{x})^2} \right )^\Delta
 \label{(2.12)}
\ee
with the following normalization
\be \label{(2.13)}
C_\Delta ={\Gamma (\Delta )\over \pi ^\d \Gamma (\Delta -\d)}\,.
%\qquad {\rm for } \qquad \Delta >\d,
%\qquad \qquad C_{\d} = {\Gamma (\d) \over 2 \pi ^\d}
\ee

Bulk--to--bulk propagators for  scalar fields of dimension
$\Delta$, with mass $m^2 = \Delta (\Delta -d)$,  were
derived in \cite{prop}. They can be expressed as hypergeometric
functions in several equivalent ways. The expression which appears best suited
for the integrals which occur in exchange diagrams
\cite{liu,dhfscalar}
 is to use a hypergeometric function whose argument is $\xi^2$ where
\be
\xi \equiv {1 \over 1+u} = { 2 z_0 w_0 \over \left( z^2_0 + w^2_0 + (\vec{z} -
  \vec{w})^2 \right)} \, .
 \label{(2.8bis)}
\ee
The propagator is then given by
\be
G_\Delta (u)
 =
   2^\Delta \tilde C _\Delta \xi ^\Delta 
   F \big ({\Delta \over 2}, {\Delta \over 2} + \half; \Delta -{d\over 2} +1;\xi^2 \big )
\, .
 \label{(2.9)}
\ee 
%{Wrong C put in, must fix}
\be \label{(2.9bis)}
\tilde C _\Delta = {\Gamma (\Delta) \Gamma (\Delta -\d +\half)
                        \over  (4\pi)^{(d+1)/2} \Gamma (2 \Delta -d+1)}
\ee
The propagator for massless scalars, with $\Delta =d$, is relevant for the
graviton. 
When $\Delta=d$ is an even integer, the hypergeometric expression
(\ref{(2.9)})
 can be
rewritten \cite{january} in terms of elementary functions.
 In particular, for $d=4$, we have
\be
G_4 (u)  = -{1 \over 8\pi ^2} \left\{ {2(1+u) \over \sqrt {u(2+u)}}
          - {1+u \over \sqrt {u(2+u)}^3} -2 \right\} 
 \label{(2.11)}
\ee

The graviton propagator \cite{january}
 can be expressed as a superposition of 5
independent fourth rank bi--tensors, of which 2 are gauge independent and
3 are gauge artifacts. The gauge terms represent pure diffeomorphisms,
and their contribution to the integrals in the exchange diagram vanishes
because the stress tensor is conserved. The physical part of the propagator 
involves two scalar functions $G(u)$ and $H(u)$, and is given by 
\bea
G_{\mu \nu \mu' \nu'} (z,w)
& =& 
\bigl (\dmu \dmup u \ \dnu \dnup u  + \dmu \dnup u \ \dnu \dmup u \bigr ) \ G(u)
+ g_{\mu \nu} \ g_{\mu ' \nu '} \ H(u)
 \label{(2.14)}
\eea
The function $G(u)$ is equal to the massless scalar propagator $G_d$. 

A representation of 
$H(u)$ as a hypergeometric function was given in \cite{january}.
 It was
also expressed in terms of $G(u)$ and its first integral $\bar
G(u)$, defined by $\bar G(u)'=G(u)$ and the boundary condition
$\bar G(\infty)=0$, which is a more useful form, given by
%as well,
%$$
%\bar G(u) = {2 \tilde C_d \over d-1} (2u^{-1})^{d -1} F(d-1,\d +\half;
%d+1;-2u^{-1})\, .
% \label{(2.16)
%$$
%The expression for $H(u)$ is then
\be
-(d-1) H(u)  = 2(1+u)^2 G(u) + 2 (d-2) (1+u) \bar G(u)  \, .
 \label{(2.17)}
\ee
Again, when $d$ is even, $H(u)$ admits an elementary expression;
in particular, when $d=4$, we have
\be
H(u)  = -{1 \over 12\pi ^2}\{-6 (1+u)^4 +  9(1+u)^2 -2\} {1+u
\over
(u(2+u))^{\frac{3}{2} }} -{1 \over 2 \pi ^2} (1+u)^2\, . 
 \label{(2.18)}
\ee

\subsection{Structure of the graviton exchange amplitude}

The graviton exchange amplitude associated with the Witten diagram of 
Figure 2t
 is given by
\be
I_{{\rm grav}}= \frac{1}{4} \,\int d z \sqrt g \int d w \sqrt g \ T_{13} ^{\mu \nu} (z) \
G_{\mu
\nu\mu'\nu'} (z,w) \ T_{24} ^{\mu ' \nu'} (w)
\label{(3.1)}
\ee
where $G_{\mu \nu\mu' \nu'}$ is the graviton propagator  (\ref{(2.14)}). 
The
vertex factor $T_{13} ^{\mu \nu}(z)$ is given by
\bea\label{(3.2)}
T_{13} ^{\mu \nu} (z) 
&= & ~ D^\mu K_\Delta (z, x_1) D^\nu K_\Delta (z, x_3) 
    + D^\nu K_\Delta (z, x_1) D^\mu K_\Delta (z, x_3)
\\
&& - g^{\mu \nu} \bigl [ \partial _\rho K_\Delta (z,x_1) D^\rho K_\Delta 
(z,x_3)
+ m^2 K_\Delta (z,x_1) K_\Delta (z,x_3) \bigr ]\, . 
 \nonumber
\eea
The combination $T_{24} ^{\mu'\nu'} (w)$ is obtained from (\ref{(3.2)}) 
by replacing
$x_1\to x_2$, $x_3 \to x_4$, $\Delta \to \Delta '$ 
$z \to w$. The stress--energy tensor $T_{\mu \nu}$ is
conserved, $D_\mu T^{\mu \nu} _{13}= D_{\mu'}
T^{\mu'\nu'}_{24}=0$ thanks to the propagator equations 
$(\Box  - m^2) K_\Delta = (\Box  - m'^2) K_{\Delta '} =0$.

It is the high tensorial rank of the propagator and vertex factors that make
this amplitude more difficult than previously studied
exchanges. The calculation is made tractable by splitting the amplitude
into several terms and using partial integration of derivatives. There are
several ways to organize this process, and what we have done and will present
are complete calculations by two different methods which are then compared and
shown to give identical results for the special case $d=\Delta=\Delta'=4$, 
{\it i.e}.
axions and dilatons in the type IIB theory. The two methods are 
separately presented in Sections 4 and 5.

\setcounter{equation}{0}
\section{The graviton exchange graph for $\Delta= \Delta'=d=4$}
%%%%%%%%%%%%%%%%%%%%%%%%%%%%%%%%%%figure%%%%%%%%%%%%%%%%%%%%%%%%%%%%%%%%%%%

\begin{figure} 
\begin{center} \PSbox{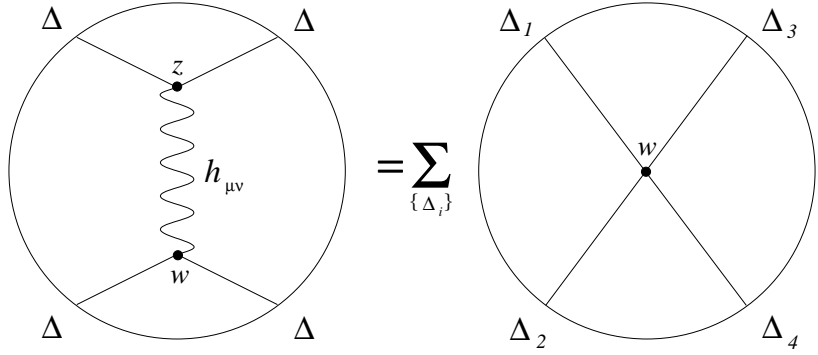 hscale=70 vscale=70}{4.6in}{1.6in} \end{center}
%\begin{center} \PSbox{scalquart.eps hscale=70 vscale=70}{4.6in}{1.7in} \end{center}
\caption{Reduction of graviton exchange to quartic graphs.}
\end{figure}

%%%%%%%%%%%%%%%%%%%%%%%%%%%%%%endfigure%%%%%%%%%%%%%%%%%%%%%%%%%%%%%%%%%%%
The graviton propagator  involves  non--trivial tensorial structures. 
Nevertheless, it turns out that it is possible to reduce the graviton exchange
graph 
to the sum of purely scalar amplitudes, with 
a peculiar pattern of bulk--to--bulk and bulk--to--boundary scalar propagators. We describe
this reduction in Section 4.1.

Furthermore, upon integration over one of the two bulk variables, which we carry out in Section 4.2, 
each effective scalar exchange  can be expressed
a sum of quartic graphs  with appropriate  external dimensions. 
The final answer for the graviton exchange in terms of these 
basic building blocks (see Figure 6) is given in equation (\ref{finalanswerDus}).
The quartic graphs admit asymptotic series expansion which we
describe in Section 6. 
It is also worth mentioning at this point that each quartic 
graph can be obtained by 
taking successive derivatives of a single basic function, see section A.3.

%%%%%%%%%%%%%%%%%%%%%%%%%%%%%%%%%%%% our section %%%%%%%%%%%%%%%%%%%%%%%%%%%%%%%%%%%%%%%%%%%%%

\subsection{Reduction to scalar exchanges}
We need to compute the graviton exchange amplitude (\ref{(3.1)}) for $m^2=m'^2=0$. Using
the form (\ref{(2.14)}) for the graviton propagator, we have\footnote{We introduce the notation $[dz]\equiv \sqrt{g}\,d^5z$.}
 :
\bea
I_{{\rm grav}}& =&  (C_4)^4(I_H + I_G) \\
 I_H &=&
\int [dz] [dw]
%\int d^5z \sqrt{ g(z)}d^5w \sqrt{ g(w)} 
\left[ 
\partial^{\mu}\tilde K_4(z,x_1)\partial^{\nu}\tilde K_4(z,x_3) -\frac{1}{2}  g^{\mu \nu}
\partial_{\lambda}\tilde K_4(z,x_1)\partial^{\lambda}\tilde K_4(z,x_3) \right]  \cdot
\\
&&   g_{\mu \nu} g_{\mu' \nu'}\,H(u)  
\left[ 
\partial^{\mu'}\tilde K_4(w,x_2)\partial^{\nu'}\tilde K_4(w,x_4) -\frac{1}{2}  g^{\mu' \nu'}
\partial_{\lambda'}\tilde K_4(w,x_2)\partial^{\lambda'}\tilde K_4(w,x_4) \right]  
\nonumber \\ 
 I_G &= &
\int [dz] [dw]
%\int d^5z \sqrt{ g(z)}d^5w \sqrt{ g(w)} 
\left[ 
\partial^{\mu}\tilde K_4(z,x_1)\partial^{\nu}\tilde K_4(z,x_3) -\frac{1}{2}  g^{\mu \nu}
\partial_{\lambda}\tilde K_4(z,x_1)\partial^{\lambda}\tilde K_4(z,x_3) \right]\cdot  \\
&& (\partial_{\mu} \partial_{ \mu'} u \,\partial_{\nu} \partial_{ \nu'} u
+\partial_{\mu} \partial_{ \nu'} u \,\partial_{\nu} \partial_{ \mu'} u ) \,
 G(u)\cdot \nonumber \\
&& \left[ 
\partial^{\mu'}\tilde K_4(w,x_2)\partial^{\nu'}\tilde K_4(w,x_4) -\frac{1}{2}  g^{\mu' \nu'}
\partial_{\lambda'}\tilde K_4(w,x_2)\partial^{\lambda'}\tilde K_4(w,x_4) \right] 
 \nonumber
\eea
where $C_4= \frac{6}{\pi^2}$ is the normalization factor (\ref{(2.13)}) of the
bulk--to--boundary propagator.
The tensorial structures in $I_H$ immediately trivialize:
\bea
I_H &=& \left(1-\frac{5}{2}\right)^2 \int [dz] [dw] \partial_{\mu}\tilde K_4\partial^{\mu}\tilde K_4 
\; H(u) \;
\partial_{\mu'}\tilde K_4\partial^{\mu'}\tilde K_4  \\
&=&\left(\frac{9}{4}\right) \int [dz] [dw] \tilde K_4 \tilde K_4  \; \frac{1}{4}\Box^2 H(u)\;
\tilde K_4 \tilde K_4 
\eea
where  we have used integration by parts
and the equation of motion $\Box \tilde K_4 =0$
to eliminate the derivatives on the $\tilde K$'s. 

Now we consider $I_G$, and it is useful to split into 4 parts:
\bea
I_G & = & I_G^1 + I_G^2+ I_G^3 +I_G^4 \nonumber \\
 I_G^1& =& \int [dz] [dw]\;\partial^{\mu}\tilde K_4(z,x_1)\partial^{\nu}\tilde K_4(z,x_3)\cdot  \label{I1}\\
&& (\partial_{\mu} \partial_{ \mu'} u \,\partial_{\nu} \partial_{ \nu'} u
+\partial_{\mu} \partial_{ \nu'} u \,\partial_{\nu} \partial_{ \mu'} u ) \,
 G(u)\,\partial^{\mu'}\tilde K_4(w,x_2)\partial^{\nu'}\tilde K_4(w,x_4) \nonumber \\
I_G^2& =&\int [dz] [dw]\;\partial^{\mu}\tilde K_4(z,x_1)\partial^{\nu}\tilde K_4(z,x_3)\cdot
 \\
&& (\partial_{\mu} \partial_{ \mu'} u \,\partial_{\nu} \partial_{ \nu'} u
+\partial_{\mu} \partial_{ \nu'} u \,\partial_{\nu} \partial_{ \mu'} u ) \,
 G(u) \left(-\frac{1}{2}  g^{\mu' \nu'}
\partial_{\lambda'}\tilde K_4(w,x_2)\partial^{\lambda'}\tilde K_4(w,x_4) \right) \nonumber \\
I_G^3 &=&\int [dz] [dw] \left(-\frac{1}{2}  g^{\mu \nu}
\partial_{\lambda}\tilde K_4(z,x_1)\partial^{\lambda}\tilde K_4(z,x_3) \right)\cdot\\
&& (\partial_{\mu} \partial_{ \mu'} u \,\partial_{\nu} \partial_{ \nu'} u
+\partial_{\mu} \partial_{ \nu'} u \,\partial_{\nu} \partial_{ \mu'} u ) \,
 G(u)\partial^{\mu'}\tilde K_4(w,x_2)\partial^{\nu'}\tilde K_4(w,x_4) \nonumber \\
I_G^4 &=&\int [dz] [dw] \left(-\frac{1}{2}  g^{\mu \nu} \label{I4}
\partial_{\lambda}\tilde K_4(z,x_1)\partial^{\lambda}\tilde K_4(z,x_3)\right)\cdot \\
&& (\partial_{\mu} \partial_{ \mu'} u \,\partial_{\nu} \partial_{ \nu'} u
+\partial_{\mu} \partial_{ \nu'} u \,\partial_{\nu} \partial_{ \mu'} u ) \,
 G(u)\left(-\frac{1}{2}  g^{\mu' \nu'}
\partial_{\lambda'}\tilde K_4(w,x_2)\partial^{\lambda'}\tilde K_4(w,x_4) \right) \nonumber
\eea
We wish to eliminate all the tensor indices and all the
derivatives, so that the graviton exchange is reduced to a sum of
effective scalar graphs. With this program in mind, we observe a few
pretty identities.
First:
\bea\label{KuK}
&&\partial^\mu \tilde K_\Delta(z,x_1) \,\partial_\mu\partial_{\nu'} u\,
\partial^{\nu'} \tilde K_\Delta(w,x_2)=\nonumber\\
&&\Delta^2\left[-\tilde K_\Delta(z,x_1)\tilde K_{\Delta+1}(w,x_2)\tilde K_{-1}(z,x_2)-\tilde K_{\Delta+1}(z,x_1)\tilde K_\Delta(w,x_2)\tilde K_{-1}(w,x_1)
 \right.\nonumber\\
&&\left.+2x_{12}^2 \tilde K_{\Delta+1}(z,x_1) \tilde K_{\Delta+1}(w,x_2)+(1+u)\tilde K_\Delta(z,x_1) \tilde K_\Delta(w,x_2) \right]
\eea
It is simplest to verify this identity by the methods described in
\cite{march}, where one uses conformal transformations to go to a
coordinate system where point $x_1$ is mapped to infinity and point
$x_2$ to zero. Further:
\be\label{Ku}
\tilde K_{\Delta+1}(z,x_1)\,\tilde K_{-1}(w,x_1)={1\over \Delta}\partial^\mu
\tilde K_\Delta(z,x_1)\partial_\mu u+(1+u)\tilde K_\Delta(z,x_1)
\ee
Inserting twice (\ref{Ku}) into (\ref{KuK}) we get:
\bea\label{KuK1}
&&\partial^\mu \tilde K_\Delta(z,x_1) \,\partial_\mu\partial_{\nu'} u\,
\partial^{\nu'} \tilde K_\Delta(w,x_2)=\nonumber\\
&&\Delta^2\left[-{1\over \Delta}\tilde K_\Delta(z,x_1) \,\partial^{\mu'} \tilde K_\Delta(w,x_2)\,
\partial_{\mu'} u-{1\over \Delta}\partial^\mu \tilde K_\Delta(z,x_1) \,\partial_\mu u\,
\tilde K_\Delta(w,x_2)\right.\nonumber\\
&&\left.+2x_{12}^2\, \tilde K_{\Delta+1}(z,x_1)\, \tilde K_{\Delta+1}(w,x_2) - (1+u) \tilde K_\Delta(z,x_1)\, \tilde K_\Delta(w,x_2) \right]
\eea
We now evaluate (\ref{I1}--\ref{I4}) one by one.
\subsubsection{$I_G^1$}
Writing (\ref{I1}) as 
\bea
I_G^1=
\int [dz] [dw]\;\left(\partial^{\mu}\tilde K_4(z,x_1)\,\partial_{\mu}
\partial_{ \mu'}
u\,\partial^{\mu'}\tilde K_4(w,x_2)\right)\,G(u)\times\nonumber\\
\times\left(\partial^{\nu}\tilde K_4(z,x_3)\,\partial_{\nu}
\partial_{ \nu'} u\,\partial^{\nu'}\tilde K_4(w,x_4)\right)
+\{x_1\leftrightarrow x_3\}
\eea
and inserting twice (\ref{KuK1}) for $\Delta=4$ we obtain $16+16$ terms many
of which are related by a simple symmetrization. Below we present the
manipulations performed on the inequivalent terms. We often suppress
the coordinate labels, and give the expressions with the propagators
in the following order: $(z,x_1)\,,(z,x_3)\,,(w,x_2)\,,(w,x_1)\,$
unless stated otherwise. Referring to the terms in (\ref{KuK1}) we
get:\\
{\bf I $\times$ I:}
\bea
{\rm I} \times {\rm I} & = & 4^2 \int [dz\,dw] \tilde K_4 \tilde K_4\, G
\,\partial^{\mu'}\tilde K_4\,\partial_{\mu'}u\,\partial^{\nu'}\tilde K_4\,\partial_{\nu'}u=\\
&=&4^2 \int [dz\,dw] \tilde K_4 \tilde K_4 \left[D_{\mu'}\partial_{\nu'}\int\int^u
G-g_{\mu'\nu'}(1+u)\,\int^u G\right]\times\nonumber\\&& \times\left(T^{\mu'\nu'}+{1\over
2}g^{\mu'\nu'}\partial^{\lambda'}\tilde K_4\,\partial_{\lambda'}\tilde K_4\right)\nonumber\\
&=&4^2 \int [dz\,dw] \tilde K_4 \tilde K_4\left[{1\over 4}\Box^2\int\int^u
G-{1\over 2}\Box\left\{(1+u)\int^u G\right\}\right] \tilde K_4 \tilde K_4 \nonumber\label{I11}
\eea
Here we have used\footnote{Here our convention is
that $\int^u F=\int^u_a\,F(u)\,dw,$ where $a$ is chosen to
ensure the fastest possible falloff of $\int^u F$ in the
$u\rightarrow\infty$ limit.}  $G(u)\,\partial_{\mu'}\partial_{\nu'}u= \left[D_{\mu'}\partial_{\nu'}\int\int^u
G-g_{\mu'\nu'}(1+u)\,\int^u G\right]$  thanks to (\ref{2.9c}) and 
we also used the conservation of the
stress-energy tensor integrating by parts to get the last equality.\\
{\bf I $\times$ II:}
\be
{\rm I} \times {\rm II} =  4^2 \int [dz\,dw] \tilde K_4\,\partial^\mu
\tilde K_4 \partial_\mu u\, G\,\partial^{\mu'}\tilde K_4 \partial_{\mu'} u \, \tilde K_4
\ee 
Using $\partial_\mu u\,G(u)=\partial_\mu \int^u G$ we get by
integration by parts:
\bea
{\rm I} \times {\rm II} & = & -4^2 \int [dz\,dw] \tilde K_4\,\partial^\mu
\tilde K_4 \partial_\mu u\, \int^u G\,\partial^{\mu'}\tilde K_4\, \partial_{\mu'} \tilde K_4\label{I12} \\
&&-4^2 \int [dz\,dw] \tilde K_4\,\partial^\mu
 \tilde K_4  \int^u G\,\partial_\mu\partial_{\mu'} u\,\partial^{\mu'}\tilde K_4\,  \tilde K_4
\nonumber
\eea
where we have used $\Box \tilde K_4=0$ in the bulk.
The first term in (\ref{I12}) can be easily processed to give
\be
4^2 \int [dz\,dw] \tilde K_4 \tilde K_4\:{1\over 4}\Box^2\int\int^u
G\: \tilde K_4 \tilde K_4
\ee
The second term in (\ref{I12}) is handled by inserting again the
identity (\ref{KuK1}) with $(x_1\leftrightarrow x_3)$ and going
through by now familiar manipulations. It gives
\bea
\int [dz\,dw] \tilde K_4\, \tilde K_4\left[-4^3 \Box\int\int^u G+4^4(1+u)\int^uG
\right]\tilde K_4\,\tilde K_4-\nonumber\\
-2\cdot4^4x_{32}^2\int\int [dz\,dw] \tilde K_4\, \tilde K_5\int^u
G\:\tilde K_5\,\tilde K_4
\eea
{\bf I $\times$ III:}
Upon integration by parts,
\bea
{\rm I} \times {\rm III}& =& 2\cdot 4^3x_{34}^2\int [dz\,dw] \tilde K_4\, \tilde K_5\:\int^u
G\:\partial^{\mu'}\tilde K_4\,\partial_{\mu'}\tilde K_5\\
& = & 2\cdot 4^3x_{34}^2\int [dz\,dw] \tilde K_4\, \tilde K_5\:\int^u
G \left[ \frac{1}{2} \Box \left( \tilde K_4\,\tilde K_5 \right) -\frac{m_5^2}{2}
\tilde K_4\,\tilde K_5 \right] \nonumber \\
& = & 2\cdot 4^3x_{34}^2\int[dz\,dw] \tilde K_4\, \tilde K_5\:\left(
\frac{1}{2} \Box\int^u G -\frac{5}{2} \int^u G \right)
  \tilde K_4\,\tilde K_5   \nonumber
\eea
{\bf I} $\times$ {\bf IV:}
\be
-4^3\int [dz\,dw] \tilde K_4\,\tilde K_4\:\frac{1}{2}\Box\int^u((1+u)G)\:\tilde K_4\,\tilde K_4
\ee
{\bf III $\times$ III:}
\be
4\cdot 4^4x_{12}^2x_{34}^2\int [dz\,dw] \tilde K_5\,\tilde K_5\:G\tilde K_5\,\tilde K_5
\ee
{\bf III $\times$ IV:}
\be
-2\cdot 4^4 x_{12}^2\int [dz\,dw] \tilde K_5\,\tilde K_4\:G(1+u)\tilde K_5\,\tilde K_4
\ee
{\bf IV $\times$ IV:}
\be
4^4\int [dz\,dw]\tilde K_4\,\tilde K_4\:G(1+u)^2\tilde K_4\,\tilde K_4
\ee
\subsubsection{$I^2_G,\;I^3_G$ and $I^4_G$}
Using (\ref{2.9b}), after some similar algebra we arrive at
\bea
I_G^2 = I_G^3=- \int [dz\,dw]\tilde K_4\,\tilde K_4 \: \frac{1}{4} \Box^2 \left[ G
+\frac{3}{2} (1+u) \int^u G + \frac{1}{2} u(u+2) G
\right] \tilde K_4\,\tilde K_4 \\
I_G^4 = \frac{1}{2} \int [dz\,dw]\tilde K_4\,\tilde K_4 \: \frac{1}{4} \Box^2 \left[5 G
+ u(u+2) G \right] \tilde K_4\,\tilde K_4 ~~~~~~~~~~~~~~~~~~~~~~~~~~~~~~~~~~~~~~~~~
\eea

\subsubsection{The graviton amplitude in terms of scalar exchanges}
Adding all the terms above with the appropriate symmetrizations we get
the complete graviton graph in terms of effective scalar exchanges:
\bea \label{I}
{I_{{\rm grav}}\over(C_4)^4} = \int [dz\,dw]\tilde K_4\,\tilde K_4 \left[  \Box^2\left( \frac{9}{16} H + 2
\cdot 4^2 \int \int^u G
+\frac{1}{8} G -\frac{3}{4}(1+u) \int^u G -\frac{1}{8} u (u+2) G
\right)\right. \nonumber \\ \left.
+ \Box \left( -4^4 \int^u \left( (1+u) G \right)
-\frac{1}{2} 4^3 (1+u) \int^u G -4^4 \int\int^u G)\right) \right. ~~~~~~~~~~~\nonumber
\\ \left. + 4^5 (1+u)
\int^u G +2\cdot 4^4 (1+u)^2 G \right] \tilde K_4\,\tilde K_4
~~~~~~~~~~~~~~~~~~~~~~~~~~~~~~~~~~~~~~~~ \\
+   x_{34}^2\int [dz\,dw]\tilde K_4\,\tilde K_5\:\left[ -18\cdot 4^3 \int^uG+2\cdot
4^3\Box\int^uG-2\cdot 4^4G(1+u)\right]\tilde K_4\,\tilde K_5\nonumber\\
+\left\{3\;{\rm perms}\right\} \nonumber \\
+(x_{12}^2x_{34}^2 +x_{14}^2x_{23}^2)\int
[dz\,dw]\tilde K_5\,\tilde K_5\:4\cdot 4^4G\:\tilde K_5\,\tilde K_5~~~~~~~~~~~~~~~~~~~~~~~~~~~~~~~~~~~~~~~~~~~~~~\nonumber
\eea
where the 3 permutations of the second integral are obtained by
exchanging $(x_1,x_2)\leftrightarrow (x_3,x_4)$ and
$x_1\leftrightarrow x_3.$
The formula above can be simplified by explicit application of Laplace
operator (\ref{box}) and using the equations 
obeyed by $G$ and $H$ given in Section 3.1. We get
\bea \label{effectivescalars}
 {I_{{\rm grav}}\over (C_4)^4}=\int
[dz\,dw]\tilde K_4\,\tilde K_4\:\left[(-72u^2-144u+168)\,G+168\,(u+1)\int^uG\right]\:\tilde K_4\,\tilde K_4
\nonumber\\
+   x_{34}^2\int [dz\,dw]\tilde K_4\,\tilde K_5\:\left[  -768\int^uG -256\,G(1+u)\right]\tilde K_4\,\tilde K_5\nonumber\\
+\left\{3\;{\rm perms}\right\} \nonumber \\+(x_{12}^2x_{34}^2 
+x_{14}^2x_{23}^2)\int
[dz\,dw]\tilde K_5\,\tilde K_5\:1024\,G\:\tilde K_5\,\tilde K_5\nonumber\\+10\, \int
[dw]\tilde K_4\,\tilde K_4\,\tilde K_4\,\tilde K_4-16\, x_{24}^2 \int
[dw]\tilde K_4\,\tilde K_4\,\tilde K_5\,\tilde K_5
\eea
The last two terms in this expression arise from delta functions
generated by the application of the Laplace operator\footnote{The
coordinate dependence of the $K$'s is: $(w,x_1),(w,x_3) ,(w,x_2),(w,x_4)$.}. In particular the
last term comes from:
\bea
&& \int[dz\,dw]\tilde K_4\,\tilde K_4\: \Box \delta(z,w) \:\tilde K_4\,\tilde K_4 \nonumber \\ &&= 2 \int[dw]\tilde K_4\,\tilde K_4
\:\partial_{\mu'} \tilde K_4 \partial^{\mu'}\tilde K_4  
=  2 \int[dw]\tilde K_4\,\tilde K_4
\left( 16 \tilde K_4 \tilde K_4 -32 x_{24}^2 \tilde K_5 \tilde K_5 \right) 
\eea
where in the last equality we have used (\ref{derivident}).

\subsection{Reduction to quartic graphs} 
We first observe  the identity
\be \label{intG}
\int^u G = -G_3 + (1+u) G\, ,
\ee
where $G_3$ is a scalar propagator of $m^2 = -3$, corresponding
to a boundary conformal dimension $\Delta=3$:
\be
-(\Box +3)G_3 =\delta(z,w)\,.
\ee
Using (\ref{intG}), we see that the complete graviton
graph (\ref{effectivescalars}) involves effective scalar exchanges
of the form 
\be \label{oneeffectivescalar}
I_{\Delta_1 \Delta_3 \Delta_2 \Delta_4}^{\Delta_5,p}
\equiv \int [dz\,dw]\tilde K_{\Delta_1} \,\tilde K_{\Delta_3} \: (1+u)^p G_{\Delta_5}   \:
\tilde K_{\Delta_2}\,\tilde K_{\Delta_4}\,,
\ee
  plus some quartic interactions
(last line of (\ref{effectivescalars})). 

We now proceed to derive a general formula to perform the
$z$ integration in (\ref{oneeffectivescalar}), following   
the methods developed in \cite{dhfscalar}. 
Quite remarkably, upon integration over $z$, (\ref{oneeffectivescalar})
reduces to a finite sum of effective quartic graphs, see Figure 6.

Translating $x_1 \rightarrow 0$ and performing conformal 
inversion (see \cite{march} for a detailed account),
we can write
\be \label{IR}
I_{\Delta_1 \Delta_3 \Delta_2 \Delta_4}^{\Delta_5,p}
=|x_{31}|^{-2 \Delta_3} |x_{21}|^{-2 \Delta_2} |x_{41}|^{-2 \Delta_4}
 \int [dw] R(w-x'_{31})
\tilde K_{\Delta_2}(w,x'_{21})\,\tilde K_{\Delta_4}(w,x'_{41})\, ,
\ee
where $\vec x'\equiv \vec x /x^2$  and  
\be
R^{\Delta_5,p}_{\Delta_1,\Delta_3}(w) =\int [dz] z_0^{\Delta_1} \tilde K_{\Delta_3}(z) (1+u)^p G_{\Delta_5}(u)\,.
\ee
As compared to \cite{dhfscalar} we allow the bulk propagator
to be multiplied by $(1+u)^p$
(see (3.3--3.4) in \cite{dhfscalar}).
We now use the hypergeometric series expansion (\ref{(2.9)}).
Inserting this series into the expression for $R^{\Delta_5,p}_{\Delta_1,\Delta_3}$,
we can perform the $z$ integral term by term by a standard
Feynman parameterization, and resum the resulting series. 
We get 
\bea
R^{\Delta_5,p}_{\Delta_1,\Delta_3}(w)&=&2^{\Delta_5-p+1}\tilde C_{\Delta_5}\pi^{d/2}{\Gamma[{1\over
2}(\Delta_5-p+\Delta_3-\Delta_1)]\Gamma[{1\over
2}
(\Delta_5-p+\Delta_1+\Delta_3-d)]\over \Gamma[\Delta-p]\Gamma[\Delta_3]}\nonumber
\\  &&\times\left({w_0 \over w^2}\right)^{\Delta_3}w_0^{\Delta_1-\Delta_3}
\int_0^1 d\gamma  {(1-\gamma )^{\Delta_3-1}\gamma ^{{1\over
2}(\Delta_5-p-\Delta_1-\Delta_3)-1}\over \left(\left({w_0 \over w^2}\right)+\gamma -\left({w_0 \over w^2}\right) \gamma \right)^{\Delta_1}} 
\\ &&\times\:
{}_4F_3\left({\textstyle{{\Delta_5\over 2}, {\Delta_5+1\over 2}, {\Delta_5-p+\Delta_3-\Delta_1\over 2}, {\Delta_5-p+\Delta_1+\Delta_3-d\over 2};
\Delta_5-{d\over 2}+1, {\Delta_5-p\over 2}, {\Delta_5-p+1\over 2};\gamma}} \right) \nonumber
\eea
For $p=0$ we recover equation (3.11) in \cite{dhfscalar}. 
It turns out that for the cases relevant to the graviton amplitude,
the hypergeometric function ${}_4 F_3$ is elementary and the Feynman
parameter integral can be explicitly done. The result is always a simple
binomial in $w_0$ and $w_0/w^2$. The relevant cases are:
\bea
R^{4,0}_{4,4}& =&{1\over 36}w_0^3\,\left({w_0\over w^2}\right)^3 + 
{1\over 48}w_0^2\,\left({w_0\over w^2}\right)^2 \nonumber\\
R^{4,2}_{4,4}& =& {1\over 36}w_0^3\,\left({w_0\over w^2}\right)^3 + 
{7\over 288}w_0^2\,\left({w_0\over w^2}\right)^2 + 
{1\over 48}w_0\,\left({w_0\over w^2}\right) \nonumber\\
R^{3,1}_{4,4}& =&{1\over 36}w_0^3\,\left({w_0\over w^2}\right)^3 + 
{1\over 36}w_0^2\,\left({w_0\over w^2}\right)^2 + 
{1\over 36}w_0\,\left({w_0\over w^2}\right) \\
R^{3,0}_{4,5}& =& {1\over 48}w_0^3\,\left({w_0\over w^2}\right)^4 + 
{1\over 48}w_0^2\,\left({w_0\over w^2}\right)^3 + 
{1\over 48}w_0\,\left({w_0\over w^2}\right)^2\nonumber\\
R^{4,1}_{4,5}& =& {1\over 48}w_0^3\,\left({w_0\over w^2}\right)^4 + 
{11\over 576}w_0^2\,\left({w_0\over w^2}\right)^3 + 
{1\over 64}w_0\,\left({w_0\over w^2}\right)^2 \nonumber\\
R^{4,0}_{5,5}& =& {1\over 64}w_0^4\,\left({w_0\over w^2}\right)^4 + 
{1\over 72}w_0^3\,\left({w_0\over w^2}\right)^3 + 
{1\over 96}w_0^2\,\left({w_0\over w^2}\right)^2\nonumber
\eea
We see that each term in $R^{\Delta_5,p}_{\Delta_1,\Delta_3}(w)$ 
 is a of product of bulk--to--boundary propagators. Indeed, 
$w_0^\Delta$ corresponds 
in this inverted frame to a propagator  at $\vec x' = \infty$,
likewise
$\left({w_0 / w^2}\right)^{\tilde \Delta}$
corresponds to a propagator at $\vec x' =0$. 
Inserting each such term
in the expression for $I_{\Delta_1 \Delta_3 \Delta_2 \Delta_4}^{\Delta_5,p}$
(equ. \ref{IR})), and going back from the inverted variables $\vec x'_i$
to the original variables $\vec x_i$, we recognize the integral defining
a quartic graph. For example
\bea
&&|x_{31}|^{-2 \Delta_3} |x_{21}|^{-2 \Delta_2} |x_{41}|^{-2 \Delta_4}
 \int [dw] w_0^{2 \Delta} \left({w_0 \over {(w - x'_{31})^2}}\right)^{\tilde\Delta}
\tilde K_{\Delta_2}(w,x'_{21})\,\tilde K_{\Delta_4}(w,x'_{41}) \nonumber \\
&& = \int [dw] \left({w_0 \over {(w- x_1)^2}} \right)^\Delta
\left({w_0 \over {(w- x_3)^2}} \right)^{\tilde \Delta}
\left({w_0 \over {(w- x_2)^2}} \right)^{\Delta_2}
\left({w_0 \over {(w- x_4)^2}} \right)^{ \Delta_4} \nonumber \\
&&\equiv D_{\Delta \tilde  \Delta\Delta_2 \Delta_4}(x_1,x_3,x_2,x_4) \, ,
\eea
where in the last line we have used the important  notation
for quartic graphs (see Figure 5) introduced in (\ref{AD}). 
We can finally write
the full graviton amplitude as a sum of quartic graphs:
\bea \label{IDasymmetric}
I_{{\rm grav}}& = &\left(\frac{6}{\pi^2} \right)^4
\Bigg[ 16\left({x_{12}^2\,x_{34}^2\over
x_{13}^2}+{x_{14}^2\,x_{23}^2\over x_{13}^2}-x_{24}^2\right)\,D_{4455}
+{128\over 9\,x_{13}^4}\left(x_{12}^2\,x_{34}^2+x_{14}^2 
\,x_{23}^2\right)\,D_{3355}\nonumber\\
&&+{32\over 3\,x_{13}^6}\left(x_{12}^2\,x_{34}^2+x_{14}^2 
\,x_{23}^2\right)\,D_{2255}+10\,D_{4444}+{14\over
3\,x_{13}^2}\,D_{3344} + {8\over
3\,x_{13}^4}\,D_{2244} \nonumber\\&&- {8\over
3\,x_{13}^6}\,D_{1144} - {16\over
3\,x_{13}^2}\,\left(x_{12}^2\,D_{4354}+x_{14}^2\,D_{4345}+x_{34}^2\,D_{3445}+x_{23}^2\,D_{3454}\right)
\nonumber\\&& -{32\over
9\,x_{13}^4}\,\left(x_{12}^2\,D_{3254}+x_{14}^2\,D_{3245}+x_{34}^2\,D_{2345}+x_{23}^2\,D_{2354}\right) \Bigg]\,.
\eea
The graviton  
amplitude  (\ref{(3.1)}) is  
symmetric under $x_1 \leftrightarrow x_3$ and $x_2 \leftrightarrow x_4$. These symmetries
are  explicit in 
the final expression for $I_{{\rm grav}}$, indeed some of the $D$ functions
(of the form $D_{\Delta \Delta \tilde \Delta
\tilde \Delta}$) are symmetric by themselves, while
asymmetric $D$ functions appear in all the symmetric permutations.
It turns out that thanks to the remarkable properties
of the $D$ functions (see equ. (\ref{symmetry})), 
the answer can be rewritten in terms
of $D_{\Delta \Delta \tilde \Delta \tilde \Delta}$'s alone. Introducing
the conformal invariant variable $s \equiv \frac{1}{2} \frac{x_{13}^2 x_{24}^2}{x_{12}^2 x_{34}^2 + x_{14}^2 x_{23}^2}$, we get 
\bea \label{finalanswerDus}
I_{{\rm grav}}& =&\left(\frac{6}{\pi^2} \right)^4 \Bigg[ 16\, x_{24}^2\left({1\over 2s}-1\right)\,
D_{4455}+{64\over 9}{x_{24}^2\over x_{13}^2}{1\over s}\, D_{3355}+
{16\over 3}{x_{24}^2\over x_{13}^4}{1\over s}\, D_{2255}\\&&+18\,D_{4444}-
{46\over 9\,x_{13}^2}\,D_{3344}-{40\over 9\,x_{13}^4}\,D_{2244} -{8\over 3\,x_{13}^6}\,D_{1144}\Bigg] \,.
\nonumber\eea
The graviton amplitude (\ref{(3.1)}) is, for the case $\Delta=\Delta'=4$
that we are considering, also
symmetric   under $(x_1,x_3) \leftrightarrow (x_2,x_4)$.
 Although not immediately
manifest in the expression above, this symmetry
 is actually present
thanks to the identity (\ref{symmetry}) obeyed by the $D$ functions.

\setcounter{equation}{0}
\section{General graviton exchange graph}

%%%%%%%%%%%%%%%%%%%%%%%%%%%%%%DHF%%%%%%%%%%%%%%%%%%%%%%%%%%%%%%%%%%%%%%%%%%%%%%%%%%%%%%%%%%%%%%%%%%%%%%%%%%%%%%%%%%%%%%%%%%%%%%%%%%%%%%%%%%%%%%%%%%%%%%%%%%%%

We expect that the amplitudes for graviton exchange between massive
scalars will be useful in general studies of the AdS/CFT correspondence.
As in past work \cite{dhfgauge,dhfscalar} we therefore assume initially that $d$, $\Delta$,
and $\Delta'$ are arbitrary, constrained only
by the unitarity bound $\Delta,\ \Delta '\geq d/2$. We will assume integer
values at the point where this step simplifies the calculation, and
specialize still later to the case $d=\Delta =\Delta'=4$ to present
detailed asymptotic formulas for dilatons and axions in the type IIB
supergravity. 

The first step in the evaluation of the amplitude (\ref{(3.1)}) is to split it into 
contributions from the terms
in $H(u)$ and $G(u)$ in the graviton propagator, and to split the latter into
a term  proportional to the metric $g^{\mu \nu}$ in  $T_{13} ^{\mu \nu} (z)$
of (\ref{(3.2)}) plus the remaining term, viz.
\be \label{(3.3)}
I_{{\rm grav}} = \frac{1}{4} \,A_{{\rm grav}} 
 = \frac{1}{4}\, (A^H + A^G _S + A^G _T) .
\ee
The three contributions are then given by
\bea
A^G _S &
 =& \int \! \! d z \sqrt g \int \! \! d w \sqrt g
   [\partial _\rho K (1) D^\rho K (3) 
    + m^2 K (1) K (3)](z)\ I_{\mu'\nu'}(z,w)\
T^{\mu'\nu'}_{24} (w)  \qquad
\label{(3.4a)} \\
A^G _T &
 =&2\int \! \! d z \sqrt g \int \! \! d w \sqrt g
   \dmu K(1) \dnu K(3)\ D^\mu \dmup u D^\nu \dnup u \ G(u) \
       T^{\mu' \nu'} _{24} (w) + (1 \leftrightarrow 3)   
 \label{(3.4b)} \\
A^H & 
 =& \int \! \! d z \sqrt g \int \! \! d w \sqrt g \
    g\cdot T_{13} (z)\ H(u)\ g\cdot T_{24} (w) 
 \label{(3.4c)} 
\eea
where we use the abbreviation $g\cdot T = g_{\mu \nu} T^{\mu \nu}$, and
\bea
I_{\mu' \nu'} & 
  \equiv & -g_{\mu \nu} G(u) [D^\mu \dmup u D^\nu \dnup u 
                    + D^\mu \dnup u D^\nu \dmup u] 
\nonumber\\
 & = & -2G(u) (g_{\mu' \nu'} +  \dmup u \dnup u)\, . 
 \label{(3.5)}
\eea
where (\ref{2.9e}) is used to obtain the second line in (\ref{(3.5)}). The symmetrization
in $1 \leftrightarrow 3$ in (\ref{(3.4b)}) will be useful for later steps.

The $w$-integral in $A^S_G$ that involves the tensor $\dmup u \dnup u$ of (\ref{(3.5)})
may be simplified by using $\dmup u G(u) = \dmup \bar G(u)$,
integrating by parts in $w$ and using the covariant conservation of $T_{24}$, 
\bea
\int dw\sqrt g G(u)  \dmup u \dnup u T_{24}^{\mu'\nu'}(w) 
 & =& -  \int dw\sqrt g \ \bar G(u) \ D_{\mu'} \dnup u \
T_{24}^{\mu'\nu'}(w)
\nonumber\\
 & = &-  \int dw\sqrt g \ (1+u) \bar G(u) \ 
g\cdot T_{24}(w)
 \label{(3.6)}
\eea 
Putting together this rearrangement of the $A^G _S$ part, we have 
\bea
A^G _S  
& =& \int \! \! d z \sqrt g \int \! \! d w \sqrt g
   [\partial _\rho K (1) D^\rho K (3) 
    + m^2 K (1) K (3)](z)\nonumber\\
&& \times \bigl \{ -2G(u) +2(1+u) \bar G(u) \bigr \} \
g\cdot T_{24} (w)
 \label{(3.7)}
\eea
Next, we use the propagator equations $(\Box -m^2)K(1) = (\Box -m^2)K(3)=0$
to obtain the following identity
\be
[\partial _\rho K (1) D^\rho K (3) + m^2 K (1) K (3)](z)
=
\half \Box _z \bigl \{ K(1) K(3) \}(z)
 \label{(3.8)}
\ee
Substituting this identity into $A^G_S$, integrating by parts the operator
$\Box _z$, neglecting vanishing boundary terms and using (\ref{box}), we find
\be
\Box_z \bigl \{ (1+u) \bar G(u) \bigr \}
= 
 -2 G(u) + 4(1+u)^2 G(u) + 2d(1+u) \bar G(u)
 \label{(3.9)}
\ee
which then gives
\bea
A^G _S  
 &=& \int \! \! d z \sqrt g  \int \! \! d w \sqrt g 
    K(1) K(3)   \bigl \{ -\Box _z G(u)  -2G(u) 
 + 4(1+u)^2 G(u)  \nonumber\\ 
                 &&  +2d(1+u) \bar G(u) \bigr \}
\ g\cdot T_{24} (w)\label{(3.10)}
\eea
Before simplifying the $g\cdot T_{24}$ factor in the integrand, we first
treat  $A^G_T$ and $A^H$ in a similar manner. For $A^H$, we use again (\ref{(3.8)})
to simplify the $z$-integration and to cast it in the following form
\be
A^H = \int \! \! d z \sqrt g \int \! \! d w \sqrt g \
    K(1) K(3) \bigl \{ -\half (d-1) \Box _z H(u) - 2 m^2 H(u) \bigr \} \
g\cdot T_{24} (w) 
 \label{(3.11)}
\ee

To simplify $A^G_T$, we begin with partial integration of $\dnu$ in
the $z$-integral in (\ref{(3.4b)}), and split $A^G_T$ as follows
\be A^G_T = -2 A^G_{T1}
- 2 A^G _{T2} \label{(3.12)} \ee where \bea A^G_{T1} & =& \int \!\! dz
\sqrt g \int \!\! dw \sqrt g \dmu \bigl \{ K(1) K(3) \bigr \} D_\nu
\bigl [D^\mu \dmup u D^\nu \dnup u G(u) \bigr ] \ T_{24} ^{\mu'\nu'}
(w) \label{(3.13a)} \\ A^G_{T2} & =& \int \!\! dz \sqrt g \int \!\! dw
\sqrt g D_\nu \dmu K(1) K(3) \big [D^\mu \dmup u D^\nu \dnup u G(u)
\big ] T_{24}^{\mu'\nu'} (w) + (1 \leftrightarrow 3)\, . \qquad \quad
\label{(3.13b)} \eea Now, $A^G_{T1}$ may be simplified by working out
the tensor algebra using (\ref{2.9a}--\ref{2.9e}) and again $\dnup u G(u) = \dnup
\bar G(u)$ to obtain \bea D_\nu \bigl [D^\mu \dmup u D^\nu \dnup u
G(u) \bigr ] &= &~ D_{\mu'} \bigl (\ \dots \bigr ) + D_{\nu '} \bigl
(\ \dots \bigr ) - D^\mu u \ g_{\mu'\nu'} J(u) \nonumber\\ J(u)& = & ~
(1+u)G(u) + (d+1) \bar G(u) \label{(3.14)} \eea The terms with
$D_{\mu'}$ and $D_{\nu'}$ cancel by partial integration in
(\ref{(3.13a)}) by conservation of $T_{24}$
%By integrations by part in $D_{\mu'}$ and $D_{\nu'}$ inside the expression for
%$A^G_{T1}$, and using the covariant conservation of $T_{24}$, the first and
%second terms respectively on the right hand side of the first line of (\ref{(3.14)})
%cancel out. 
Finally, integrating by parts once more in $\dmu$ and using   
$D^\mu u \ J(u) = D^\mu \int ^u J$, we get the following simple result for
$A^G_{T1}$,
\bea \label{(3.15)}
A^G_{T1} & 
 =& -\int \!\! dz \sqrt g \int \!\! dw \sqrt g
   \dmu \{ K(1) K(3)\} D^\mu u\ J(u) \ g\cdot T_{24} (w) \\
& 
 = &\int \!\! dz \sqrt g \int \!\! dw \sqrt g
 K(1) K(3) \bigl \{ u(2+u) J'(u) + (d+1)(1+u) J(u) \bigr \} \ g\cdot T_{24} (w)
 \nonumber
\eea

It is more difficult to deal with $A^G_{T2}$
To simplify the integral representation  in (\ref{(3.13b)}), it is very
convenient to set $x_1=0$ in the first term and then perform an inversion 
transformation of the integral (in $z$ and $w$) as explained in \cite{march}. 
The symmetric step in $1 \leftrightarrow 3$ is done later.
%(3.13b) -- leaving the symmetrization in $1 \leftrightarrow 3$ until 
%later -- and
%then to send $x_1'\to \infty$. 
It is now easy to evaluate the double covariant
derivative of the inverted bulk-to-boundary propagator 
$K(1')= C_\Delta z_0^\Delta$,
\be
D_\nu \dmu K(1') 
= - \Delta K(1') g_{\mu \nu} 
  + \Delta (\Delta +1) K(1') z_0 ^2 g _{\mu 0} g _{\nu 0}
 \label{(3.16)}
\ee

The contribution of the first term in (\ref{(3.16)}) is proportional to the metric
$g_{\mu \nu}$,  and may be treated by the same technique  used for  $A^G_{S}$.
It acquires an ``effective scalar propagator'' proportional to the term
in $\{\dots\}$ in (\ref{(3.7)}). We thus find for this contribution to $A^G_{T2}$ the
term
%Since we will group all similar contributions together 
%we now reexpress this term in the original coordinates $x_i$. This term is
%manifestly symmetric
%under $1\leftrightarrow 3$, so that symmetrization merely produces an overall
%factor of 2. We thus find for this contribution to $A^G_{T2}$ the term
\be
-  |x_{21}'|^{2 \Delta '} |x_{31}'|^{2 \Delta } |x_{41}'|^{2 \Delta '}
\Delta \int \!\! dz \sqrt g \int \!\! dw \sqrt g K(1') K(3')
\bigl \{ G(u) - (1+u) \bar G(u) \bigr \} \ g\cdot T_{24} (w)
+ (1 \leftrightarrow 3)\, . \qquad \quad
 \label{(3.17)}
\ee
Note that the prefactor contains the scale factors from the inversion.

%%%%%up to here on Sunday 21 Feb, but changed on Monday!
%%%%%
The integral of the second term in (\ref{(3.16)}) contains the factor.
\be
z_0 ^2 g_{\mu 0} g_{\nu 0} D^\mu \dmup u \ D^\nu \dnup u
=
 \bigl ( z_0 g_{\mu'0'} + \dmup u \bigr )
            \bigl ( z_0 g_{\nu'0'} + \dnup u \bigr )
 \label{(3.18)}
\ee
Integration in $w$ against $G(u)T_{24}^{\mu'\nu'}(w)$ gives rise to three types
of terms
\bea
&&\int \!\! dw \sqrt g  \bigl ( z_0 g_{\mu'0'} + \dmup u \bigr )
            \bigl ( z_0 g_{\nu'0'} + \dnup u \bigr )
          G(u) T_{24} ^{\mu'\nu'}(w) \nonumber\\
&& =
z_0 ^2 \int \!\! dw \sqrt g \ G(u) T_{24} (w)_{0'0'} 
+2 z_0 \int \!\! dw \sqrt g \ g_{\mu' 0'} \dnup u \ G(u)\ T_{24} ^{\mu'\nu'} (w)
\nonumber\\ 
&&~~~  
- \int \!\! dw \sqrt g \ (1+u) \bar G(u) \ g\cdot T_{24} (w).
 \label{(3.19)}
\eea
The second integral on the right hand side may be further simplified by using
once more $\dnup u G(u) = \dnup \bar G(u)$, integrating by parts, using
conservation of $T_{24}$ and being careful to taking into account the fact that
the integral is the $0'$ component of a vector instead of a scalar. Thus there
is a non-vanishing contribution of Christoffel symbols, which gives
\be
\int \!\! dw \sqrt g \ g_{\mu' 0'} \dnup u \ G(u)\ T_{24} ^{\mu'\nu'} (w)
=
\int dw \sqrt g \ {1 \over w_0} \bar G(u) \ g\cdot T_{24}(w)
 \label{(3.20)}
\ee
We now combine (\ref{(3.17)}),(\ref{(3.19)}) and (\ref{(3.20)}) to write an expression for
$   A^G _{T2}$, namely 
\bea
A^G _{T2}& = & 
 |x_{21}'|^{2\Delta '} |x_{31}'|^{2\Delta} |x_{41}'|^{2\Delta '}\int \!\! dz 
\sqrt g \int \!\! dw \sqrt g K(1') K(3') \nonumber\\
&& \bigg\{\bigg [-\Delta G(u) -  \Delta ^2 (1+u) \bar G(u)  +
\Delta (\Delta +1){2z_0\over w_0}\bar G(u) \bigg] 
\ g\cdot T_{24}(w)
\nonumber\\
&& + \Delta (\Delta +1) 
 \ z_0^2 G(u) T_{24}(w)_{0'0'} \bigg\}
\nonumber\\
&&+ (1 \leftrightarrow 3 )
 \label{(3.21)}
\eea

\subsection{Final simplified form}

We are now in a position to assemble all contributions to the graviton exchange
diagram by combining  results for $A^H$, $A^G_S$, $A^G _{T1}$ and 
$A^G_{T2}$. The $z$-integrals are easiest to carry out after inversion, so we
apply inversion to all contributions and rewrite $A_{{\rm grav}}$ with a universal 
conformal factor extracted, viz. 
\be
A_{{\rm grav}} = |x_{21}'|^{2 \Delta '} |x_{31}'|^{2 \Delta } |x_{41}'|^{2 \Delta '} (B^{tt} + B^{dd}
+B^{00}) + (1 \leftrightarrow 3)
 \label{(3.22)}
\ee
where the reduced amplitudes $B$ are given by
\bea
B^{tt}&= & 
 \int \!\! dz \sqrt g \int \!\! dw \sqrt g K(1') K(3') 
 P(u) \ g\cdot T_{24}(w)
 \label{(3.23a)} \\
B^{dd} &= & 
 - 4\Delta (\Delta +1) 
\int \!\! dz \sqrt g \int \!\! dw \sqrt g \ {z_0  \over w_0} K(1')K(3')
\bar G(u) \ g \cdot T_{24}(w)  
 \label{(3.23b)} \\
B^{00}& = & 
 - 2 \Delta (\Delta +1) \int \!\! dz \sqrt g \int \!\! dw \sqrt g
\ z_0 ^2 \ K(1') K(3') \ G(u) \ T_{24} (w) _{0'0'}
 \label{(3.23c)}  
\eea
The function $P(u)$ is gotten by combining all contributions
involving $g\cdot T_{24}$ (except that from  $A^G _{T2}$)  and is given by
\bea
P(u) & =&
-\half \Box _z G -{1\over 4} (d-1) \Box _z H -G +2(1+u)^2 G +d(1+u)
\bar G(u) -m^2 H 
\nonumber\\
&& 
-u(2+u) J' -(d+1) (1+u) J +2 \Delta G +2 \Delta ^2 (1+u) \bar G(u)
 \label{(3.24)}
\eea
The relation between $H(u)$ and $G(u)$ was given in (\ref{(2.17)}) and may be used to
further simplify the form of $P(u)$. While both $\Box _z G$ and $\Box _z
H$ have a term proportional to $\delta (z,w)$, the relative coefficients of
both terms are such that this $\delta$-functions cancels out of the full
$P(u)$, and we are left with 
\bea \label{(3.25)}
P(u) 
& = &2\Delta G -2u(2+u) G + 2(\Delta ^2 -d-1) (1+u) \bar G(u)
-m^2 H(u) \nonumber\\
& = &2 \bigl \{\Delta +1  + {m^2-d+1\over d-1}   (1+u)^2 \bigr \}G(u) 
+2\bigl \{\Delta ^2 -d-1+ {m^2 (d-2) \over d-1} \bigr \}(1+u) \bar G(u) 
 \nonumber
\eea
Finally, the expression $T_{24}(w)_{0'0'}$ may be worked out explicitly,
\bea
T_{24}(w)_{0'0'}
&=&
(\Delta ')^2 K_{\Delta '} (2') K_{\Delta '} (4')
\biggl \{ 
  \biggl ( 1-{(m')^2 \over (\Delta ')^2}\biggr ) {1 \over w_0^2}
  -{4 \over (w-x'_2)^2}
\nonumber\\ && - {4 \over (w-x'_4)^2}
  + {8w_0 ^2 + 2 (x'_2 - x'_4)^2 \over (w-x'_2)^2 (w-x'_4)^2} 
\biggr \} \,,
\label{(3.26)}
\eea
and we can use an identity similar to (\ref{(3.8)}) to obtain a 
covariant expression for $g \cdot T_{24}(w)$, namely
\be
g\cdot T_{24}(w)
 = \bigl ( -\half (d-1) \Box _w - 2 m'^2 \bigr ) \bigl \{ K(2') K(4') \bigr \}
 \label{(3.27)}\,.
\ee

\subsection{ General integrals over interaction points}

We shall use the following strategy for the calculation of the integrals over
the interaction points $z$ and $w$ in the reduced amplitudes of (\ref{(3.23a)}--\ref{(3.23c)}). First,
we shift both $z$ and $w$ by $x'_{31}$; by translation invariance,  the integrals depend only upon the new variables
$x\equiv x_{41}'-x_{31}'$ and $y\equiv x_{21}'-x_{31}'$. The $z$-integrations then only
depend upon the variable $w$, and may be carried out explicitly in terms of
elementary functions by methods similar to the ones used in \cite{dhfgauge} and 
\cite{dhfscalar}. Only after the $z$-integrals are carried out are the explicit forms of $g\cdot
T_{24}$ and $T_{24}(w)_{0'0'}$ required and used. The remaining
$w$-integrals may be recast as integral representations that admit
simple asymptotic expansions.

To prepare for the $z$--integrations, we note that $P(u)$ in (\ref{(3.23a)}))
and (\ref{(3.25)}) involves the invariant function $G(u)$ and 
its first integral $\bar G(u)$, and the same functions appear
in (\ref{(3.23b)},\ref{(3.23c)}). To apply the methods of \cite{dhfgauge} and \cite{dhfscalar}
 we need the series expansions of $G(u)$ and $\bar G (u)$ in the variable
$\xi$ of (\ref{(2.8bis)}). For $G(u)$ this is
just the hypergeometric series for $G_d(u)$ in (\ref{(2.9)}) and we 
obtain the series for $\bar G(u)$ by direct integration.
These expansions are given by 
\bea
G(u) & 
=& \half C_G\sum _{k=0} ^\infty 
   {\Gamma (k+\d+\half) \over \Gamma (\d +\half) \ k!}\ {1 \over k+\d}\ \xi
^{2k+d} \nonumber\\
\bar G(u) &=& -{1 \over 4}C_G \sum _{k=0} ^\infty 
   {\Gamma (k+\d-\half) \over \Gamma (\d +\half) \ k!}\ {1 \over k+\d}\ \xi
^{2k+d -1} 
 \label{(4.2)}
\eea
These series expansions are uniformly convergent inside any disc $|\xi|<1$. The
normalization constant may be read off from (\ref{(2.9bis)}) and (\ref{(2.13)}) for $\Delta =d$,
and we find $C_G = 2^d \ d \ \tilde C_d$.

There are five independent $z$-integrals required to evaluate the graviton
exchange amplitudes. They are as follows,
\bea
Z_1(w) & =& \int dz\sqrt g \ K(1') K(3') G(u)
 \label{(4.3a)} \\
Z_2(w) & =& \int dz\sqrt g \ K(1') K(3') (1+u)^2 G(u)
  \label{(4.3b)} \\
Z_3(w) & =& \int dz\sqrt g \ K(1') K(3') (1+u) \bar G(u)
  \label{(4.3c)} \\
Z_4(w) & =& \int dz\sqrt g \ K(1') K(3') z_0 w_0 ^{-1} \bar G(u)
  \label{(4.3d)} \\
Z_5(w) & =& \int dz\sqrt g \ K(1') K(3') z_0^2 w_0^{-2} G(u)
  \label{(4.3e)} 
\eea
In terms of these integrals, the original amplitudes are given by
\bea
B^{tt} & =&
 \int dw \sqrt g \ \bigl \{ 
   2(\Delta +1) Z_1(w) + 2{m^2 -d+1 \over d-1} Z_2(w) \nonumber\\
&&+2(\Delta ^2 -d-1 + {m^2 (d-2)\over d-1} )Z_3(w) \bigr \}
g\cdot T_{24}(w) \label{(4.4a)} \\
B^{dd} & =&
  \int dw \sqrt g \ \bigl \{
   -4\Delta (\Delta +1) Z_4(w) \bigr \} g\cdot T_{24}(w)
\label{(4.4b)} \\
B^{00} & =&
 \int dw \sqrt g \ \bigl \{ - 2\Delta (\Delta +1) Z_5(w) \bigr \} 
  w_0 ^2 T_{24}(w)_{0'0'}
 \label{(4.4c)} 
\eea
It remains to evaluate the $z$-integrals.

\subsubsection{ Performing the $z$-integrals}

 The $z$-integrations are carried out
term by term on the series expansions of (\ref{(4.2)}), and all the integrals we need in
(\ref{(3.23a)}--\ref{(3.23c)}) are of the following form (with $2a,\ 2b=0,1$ or 2)
\bea
&&\int _0 ^\infty \!\! dz_0  \int _{{\bf R}^d} \!\! d^d\vec{z} 
\ {z_0 ^{2\Delta +2a -d-1} \over ( z_0^2 +\vec{z}^2)^\Delta}
\biggl ( {2z_0 w_0 \over z_0^2 +w_0^2 +(\vec{z}-\vec{w})^2 } \biggr )^{2k+d-2b}
\nonumber\\
&& =
\pi ^\d {\Gamma(\half) \Gamma (\Delta +k +a-b) \Gamma (k+\d -a-b)
\over  \Gamma (\Delta ) \Gamma (k+\d-b) \Gamma (k+\d-b+\half)} \ w_0^{2a} \nonumber\\
&& ~~~\times
\int _0 ^1 d\alpha \ \alpha ^{2a-1} (1-\alpha )^{\Delta -1}
\biggl ( {\alpha w_0^2 \over \alpha w_0^2 + (1-\alpha )w^2} \biggr )^{k+\d-a-b}
 \label{(4.5)}
\eea
In the integrals $Z_j(w)$ of (\ref{(4.3a)}--\ref{(4.3e)}), the values taken by $(a,b)$ are $(0,0)$,
$(0,1)$, $(0,1)$, $(\half, \half)$ and $(1,0)$ for $j=1,2,3,4,5$ respectively.
The calculation of the $z$-integrals is slightly involved, but is essentially
the same for each of the $Z_j$-integrals. Here, we shall present in detail
only the calculation for $Z_1$, and restrict to presenting the final results
for the remaining 4 integrals.

To compute $Z_1(w)$, we use the expansion of (\ref{(4.2)}) for the function $G(u)$ and
integrate term by term in $z$ using the integral formula of (\ref{(4.5)}), here with
$a=b=0$. Assembling these results, we notice that the factors $\Gamma
(k+\d+\half)$ and $\Gamma (k+\d)$ cancel between numerators and denominators.
Also, interchanging the order of the $\alpha$-integration of (\ref{(4.5)}) and the
$k$-sum of (\ref{(4.2)}),  we are left with the following result
\bea
Z_1(w) & = &
 {\pi ^\d \Gamma (\half) \over 2\Gamma (\d +\half)} C_G C_\Delta ^2 
\int ^1 _0 {d\alpha
\over
\alpha} (1-\alpha )^{\Delta -1} f_{\Delta;\d} \biggl ({\alpha w_0^2 \over \alpha
w_0^2 +(1-\alpha )w^2}\biggr ) \nonumber\\
f_{\Delta;p} (\zeta) & = &\sum _{k=0} ^\infty {\Gamma (k+\Delta )
\over \Gamma (\Delta ) \ k!}\ {\zeta ^{k+p} \over k+p}
 \label{(4.6)}
\eea
Assuming that $d$ is even and $d\geq 4$ throughout, we have $p > 1$ and the
function $f_{\Delta;p}$ may be easily evaluated in terms of elementary functions.
We begin by noticing that
\be
f_{\Delta ;p}(\zeta) = \zeta ^p \biggl ( {d \over d\zeta} \biggr ) ^{p-1}
\sum _{k=0} ^\infty {\Gamma (k+\Delta ) \over \Gamma (\Delta) \Gamma (k+p+1)}
\zeta ^{k+p-1}
 \label{(4.7)}
\ee
In view of the presence of the multiple derivative operation in front, we are
free to add into the sum the terms with $k=-p+1, -p+2, \cdots ,-1$. Then, we
shift $k\to k-p$ and obtain
\be
f_{\Delta ;p}(\zeta ) = \zeta ^p \biggl ( {d \over d\zeta } \biggr ) ^{p-1}
\sum _{k=1} ^\infty {\Gamma (k+\Delta -p ) \over \Gamma (\Delta) \Gamma
(k+1)} \zeta ^{k-1}
 \label{(4.7bis)}
\ee
The infinite sum is proportional to $\zeta ^{-1} [(1-\zeta )^{-\Delta +p} -1]$
and the multiple differentiations may be carried out explicitly. The final
result is
\be
f_{\Delta ;p}(\zeta ) 
= (-)^p {\Gamma (p) \over \Gamma (\Delta)} \biggl [
 \Gamma (\Delta -p) - \sum _{\ell =0} ^{p-1}
  (-)^\ell {\Gamma (\Delta -p+\ell) \over \ell !} {\zeta ^\ell \over
(1-\zeta )^{\Delta -p+\ell}} \biggr ]
 \label{(4.8)}
\ee 
Upon substituting the value $\zeta =\alpha w_0^2/(\alpha w_0^2 +(1-\alpha )w^2)$,
and using the binomial expansion for the (positive) powers of the combination
$\alpha w_0^2 +(1-\alpha)w^2$, we find
\be
f_{\Delta ;p}(\zeta )
=
-(-)^p {\Gamma (p) \over \Gamma (\Delta)} 
\sum _{k=0}^{\Delta -2} \sum _{\ell=0} ^{p-1} (-)^\ell 
{ \Gamma (\Delta -p+\ell) \Gamma (\Delta -p+1)
  \over
  \ell ! \ \Gamma (\Delta -p+\ell-k) \Gamma (k-\ell+2)}
\biggl ( {\alpha w_0^2 \over (1-\alpha) w^2}\biggr ) ^{k+1}
 \label{(4.9)}
\ee

Remarkably, upon including the factor of $\alpha ^{-1} (1-\alpha
)^{\Delta -1}$ of the integral in (\ref{(4.6)}), the integrand is polynomial in $\alpha$
and may be carried out term by term in (\ref{(4.9)}). The final result for this
calculation as well as for that of the remaining $Z_j$ may be expressed in the
following final form
\be
Z_j(w) = 
\sum _{k=0} ^{\Delta -2} Z_j ^{(k)} \biggl (
{w_0^2 \over w^2} \biggr ) ^{k+1}
\qquad \quad j=1,\cdots ,5
 \label{(4.10)}
\ee
with the coefficients $Z_j^{(k)}$ dependent only on $\Delta$ and $d$ and given as
follows
\bea
Z_j^{(k)}& =& (-)^\d 
{\pi ^\d \Gamma (\half) \Gamma (\d) \Gamma (\Delta -\d+1) 
  \over 
  2 \Gamma (\d +\half) \Gamma (\Delta)^2} \ C_G \ C_\Delta ^2\ \hat Z_j^{(k)}
 \label{(4.11a)} \\ 
& =& (-)^\d {\Gamma (\d)\ (\Delta -\d)^2 \over 4 \pi ^d \Gamma (\Delta
-\d +1)} \ \hat Z_j ^{(k)}  ,
 \label{(4.11b)} 
\eea
with 
\bea
\hat Z_1 ^{(k)} & =&  \sum _{\ell =0} ^{\d-1} -(-)^\ell 
  {\Gamma (\Delta +\ell -\d) \Gamma (\Delta -k-1) \Gamma (k+1)
   \over
    \ell ! \ \Gamma (k-\ell +2) \Gamma (\Delta -k +\ell -\d)}
\label{(4.12a)} \\
\hat Z_2 ^{(k)} & =& \hat Z_3 ^{(k)}  (-)^\d 
  {\Gamma (\Delta -1) \Gamma (k+1)
   \over
    \Gamma (k-\d +3) }
\label{(4.12b)} \\
\hat Z_3 ^{(k)} & =&
 \half \sum _{\ell =0} ^{\d-1} (-)^\ell  
  {\Gamma (\Delta +\ell -\d-1) \Gamma (\Delta -k-1) \Gamma (k+1)
   \over
   \ell ! \ \Gamma (k-\ell +3) \Gamma (\Delta -k +\ell -\d-1)}  \label{(4.12c)} \\
\hat Z_4 ^{(k)} &=& - {\Delta -\d+1 \over (d-2) \Delta}
\sum _{\ell =0} ^{\d-2} (-)^\ell 
  { (\d-\ell-1) \Gamma (\Delta +\ell -\d) \Gamma (\Delta -k-1) \Gamma (k+2) 
   \over
    \ell ! \ \Gamma (k-\ell +3)  \Gamma (\Delta -k +\ell -\d)}
\label{(4.12d)} \\
\hat Z_5 ^{(k)} & =& {2(\Delta -\d+2)(\Delta -\d+1) \over d \Delta (\Delta +1) }
\nonumber\\
&&  \times \sum _{\ell =0} ^{\d-2} (-)^\ell 
  {(\d-\ell-1) \Gamma (\Delta +\ell -\d+1) \Gamma (\Delta -k-1) \Gamma (k+3)
   \over
    \ell ! \ \Gamma (k-\ell +3) \Gamma (\Delta -k +\ell -\d+1)} \qquad \qquad
\label{(4.12e)} 
\eea
We conclude by noticing that the relation between $Z^{(k)}_j$ and $\hat
Z^{(k)}_j$ simplifies considerably upon using the explicit forms for $C_G$ and
$C_\Delta$, as was done in (\ref{(4.11b)}). 
%for $\Delta >\d$. For the special
%case $\Delta = \frac{d}{2}$, the coefficient pf $\hat Z_j^(k)$ in
%(\ref{(4.11b)}) should be replaced by
%\be
%(-)^{\frac{d}{2}} 
%\frac{\Gamma(\frac{d}{2}) \Gamma(\frac{d}{2} +1)}{16 \pi^d}\,.
%\ee

\subsubsection{ Reduction to $w$-integrals}

Our purpose here is to express the $w$--integrals in $B^{tt}$, $B^{dd}$,
and $B^{00}$ of (\ref{(4.4a)}--\ref{(4.4c)}) in terms of the following standard
integral 
\be
W_k^{\Delta'} (a,b) 
\equiv 
\int dw \sqrt g \ { w_0 ^{2\Delta '+2a +2k} \over w^{2k}}
{1 \over (w-x)^{2\Delta'} } {1 \over (w-y)^{2\Delta '+2b}}\,.
 \label{(4.13)}
\ee
We also use $\tilde W_k^{\Delta'}(a,b)$ which represents $W_k^{\Delta'}(a,b)$ with $x
\leftrightarrow y$.
Introducing the constants
\be
Z^{(k)} =
 2(\Delta +1) Z_1 ^{(k)} + 2{m^2 -d+1\over d-1} Z_2 ^{(k)}
 +2(\Delta ^2 -d-1 + {m^2 (d-2) \over d-1})Z_3^{(k)}
-4 \Delta (\Delta +1) Z_4^{(k)}
 \label{(4.15)}
\ee
we find the following  expression for $B^{tt}+B^{dd}$, after partial integration
of $\Box_w$,
\be
B^{tt}+B^{dd}
 = \sum _{k=0} ^{\Delta -2}  Z^{(k)} \int dw \sqrt g \ 
\biggl ( -\half (d-1) \Box _w -2 m'^2 \biggr ) \biggl ({w_0 ^2 \over w^2 }
\biggr ) ^{k+1} K(2') K(4')
 \label{(4.16)}
\ee
The action of the Laplace operator on the various powers of $w_0^2/w^2$ is easily
evaluated with the help of the following formula
\be
\Box _w \biggl ( {w_0 ^2 \over w^2 }\biggr ) ^k
=
2k(2k-d) \biggl ( {w_0 ^2 \over w^2 }\biggr ) ^k
-4k^2 \biggl ( {w_0 ^2 \over w^2 }\biggr ) ^{k+1}
 \label{(4.17)}
\ee
and we obtain the following expression for the amplitude in terms of $W$
functions
\bea
B^{tt}+B^{dd}
 &=& c_{\Delta'}^2  \sum _{k=0} ^{\Delta -2}  Z^{(k)} \biggl [
\bigl \{ -(d-1)(k+1)(2k+2-d) -2m'^2 \bigr \} W^{\Delta '}_{k+1}(0,0)
\nonumber\\ && 
+ 2(d-1)(k+1)^2 W_{k+2}^{\Delta '}(0,0) \biggr ]
 \label{(4.18result)}
\eea
Proceeding analogously for the contribution of $B^{00}$ with the help of (\ref{(3.26)})
and (\ref{(4.4c)}), we find
\bea \label{(4.19)}
B^{00} &=& -2 \Delta (\Delta +1)  (\Delta ')^2 c_{\Delta'}^2 \sum _{k=0} ^{\Delta-2}
Z^{(k)} _5 \biggl \{ \bigl (1 - {m'^2 \over \Delta '^2} \bigr ) W_{k+1}^{\Delta
'}(0,0) -4 W_{k+1}^{\Delta '} (1,1) -4 \tilde W_{k+1}^{\Delta '} (1,1) 
\nonumber\\
&& + 8 W_{k+1}^{\Delta
'+1} (0,0) +2(x-y)^2 W_{k+1} ^{\Delta '+1}(0,0)  \biggr \}
\eea 
As in the special case $\Delta = \Delta' =d=4$
already discussed in Section 3, we 
recognize that the general graviton exchange
amplitude is a finite sum of quartic graphs.
In fact, each $W_k^{\Delta'}(a,b)$ is the amplitude
of a 4--point contact diagram evaluated in the inverted coordinates
(with appropriate inversion prefactors omitted). The scale dimension
of the external propagators are $\Delta_1 = k+2a-b$,
$\Delta_3=k$, $\Delta_2 = \Delta'+b$ and $\Delta_4 = \Delta'$
(see equ.(\ref{DW})).

\subsection{Graviton exchange graph for  $d=\Delta =\Delta'=4$}

For $\Delta = \Delta '=d=4$, the masses of the scalars vanish $m=m'=0$, and the
$k$ and $\ell$-sums in the results for the $z$-integral functions $I_j$
truncate after just a few terms. We need the $z$-integral functions $Z_j(w)$,
$j=1,\cdots,5$, which may be read off from 
(\ref{(4.10)}) and (\ref{(4.12a)}--\ref{(4.12e)})  with $\Delta
=d=4$,
\bea
Z_1(w) & 
 =& {1 \over 2 \pi ^4} \bigl ( \qquad \quad \ 
       +{3 \over 2} {w_0^4 \over w^4}
       +2 {w_0^6 \over w^6} \bigr ) 
 \label{(5.1a)} \\
Z_2(w) & 
 =& {1 \over 2 \pi ^4} \bigl (
       {3 \over 2} {w_0^2 \over w^2}
       +{7 \over 4} {w_0^4 \over w^4}
       +2 {w_0^6 \over w^6} \bigr ) 
 \label{(5.1b)} \\
Z_3(w) & 
 =& {1 \over 2 \pi ^4} \bigl (
       - \half {w_0^2 \over w^2} 
       -{1 \over 4} {w_0^4 \over w^4}
       \bigr ) 
 \label{(5.1c)} \\
Z_4(w) & 
 =& {1 \over 2 \pi ^4} \bigl ( 
       -{3 \over 8} {w_0^2 \over w^2}
       - {1 \over 8} {w_0^4 \over w^4} \bigr ) 
 \label{(5.1d)} \\
Z_5(w) & 
 =& {3 \over 10 \pi ^4} \bigl ( 
       + {w_0^2 \over w^2}\ +{w_0^4 \over w^4}\
       + {w_0^6 \over w^6} \bigr ) 
 \label{(5.1e)} 
\eea
Using these integrals, the expressions for $B^{tt}+B^{dd}$ and $B^{00}$ become
quite simple and are given as follows,
\bea
B^{tt} + B^{dd} &
 =& {8  \over \pi ^4 } \int dw \sqrt g \ \biggl \{ 
{w_0^2 \over w^2}+{w_0^4 \over w^4}+{w_0^6 \over w^6} \biggr \} \
g\cdot T_{24}(w)
 \label{(5.2a)} \\
B^{00} &
 =& - {12 \over \pi ^4 } \int dw \sqrt g \ \biggl \{ 
{w_0^2 \over w^2}+{w_0^4 \over w^4}+{w_0^6 \over w^6} \biggr \} \
w_0 ^2 T_{24}(w)_{0'0'}
 \label{(5.2b)} 
\eea
When $m'=0$ and $d=4$, the combination $g\cdot T_{24}$ in (\ref{(3.27)})
simplifies.
%and we have 
%\be
%g\cdot T_{24}(w) = -{3 \over 2} \Box _w \{ K(2') K(4') \}
% \label{(5.3)}
%\ee
Upon integration by parts, and making use of the differentiation
formula (\ref{(4.17)}),
we obtain the following expression 
\bea
B^{tt} + B^{dd} &
 =& {2^6 \cdot 3^3  \over \pi ^8} \int dw \sqrt g \ \biggl \{ 
{w_0^2 \over w^2}+{w_0^4 \over w^4}+{w_0^6 \over w^6} + 9 {w_0^8 \over w^8}
\biggr \} {w_0 ^8 \over (w-x)^8 (w-y)^8}
\nonumber\\
& =&
 {2^6 \cdot 3^3 \over \pi ^8} \biggl \{ W_1^4(0,0) + W_2 ^4(0,0) + W_3 ^4 (0,0)
+ 9 W_4 ^4(0,0)  \biggr \}
 \label{(5.4)}
\eea
The expression for $B^{00}$ may be obtained in an analogously, using (\ref{(3.26)})
for $m'=0$, $\Delta=4$. This directly gives
%\be
%w_0 ^2 T_{24}(w)_{0'0'}
%=
%16 K(2') K(4') \biggl \{
%1 -{4 w_0 ^2 \over (w-x)^2} - {4w_0^2 \over (w-y)^2}
%+ {8w_0 ^4 + 2 w_0 ^2 (x-y)^2 \over (w-x)^2 (w-y)^2 } \biggr \}
% \label{(5.5)}
%\ee
%and use it to obtain the full amplitude $B^{00}$ as follows
\be
B^{00} 
=
-{2^9 \cdot 3^3  \over \pi ^8} \sum _{p=1} ^3 \biggl \{
 W_p ^4 (0,0) - 4 W_p ^4(1,1) - 4 \tilde W_p^4(1,1) +8 W_p ^5 (1,0) 
+2 (x-y)^2 W_p ^5(0,0) \biggr \}
 \label{pippo}
\ee
Using the expression for $W_p ^4(1,1) + \tilde W_p^4(1,1)$ in terms of $W(0,0)$
to be derived in (\ref{(4.22)}), this formula may be recast in terms of $W(0,0)$ and $W(1,0)$ only, viz.
\be \label{(5.7)}
B^{00} 
=
-{2^9 \cdot 3^3  \over \pi ^8} \biggl [ 
3 W_4 ^4 (0,0) +  \sum _{p=1} ^3 \bigl \{
-2 W_p ^4 (0,0) +8 W_p ^5 (1,0) 
+ 2(x-y)^2 W_p ^5(0,0) \bigr \} \biggr ]
\ee
Adding the contributions of $B^{tt}+B^{dd}$ and $B^{00}$, we finally obtain the
expression for the full  $B$ in terms of $W$-functions and we have
\be
B= 
-{2^6 \cdot 3^3  \over \pi ^8} \biggl [ 
15 W_4 ^4 (0,0) +  \sum _{p=1} ^3 \bigl \{
-17 W_p ^4 (0,0) +64 W_p ^5 (1,0) 
+ 16 (x-y)^2 W_p ^5(0,0) \bigr \} \biggr ]\,.
 \label{(5.8)}
\ee
   The full graviton amplitude $I_{{\rm grav}}$ is obtained 
by multiplying $B$ by the appropriate kinematic factors  and symmetrizing under $1 \leftrightarrow 3$ (see (\ref{(3.3)}), (\ref{(3.22)})).

\subsection{Equivalence with the  result in Section 3}
We now make contact with the result obtained
in Section 4. We recall that  $W_k^{\Delta'}(a,b)$ are
just scalar quartic graphs in the inverted coordinates (with some kinematic factors
omitted), see equ.(\ref{DW}). One can easily convert (\ref{(5.8)}) and (\ref{(3.22)}) into
the notations
Section 4, and get a sum of $D$--functions.
The representation of the graviton exchange graph that is obtained in this way does not at first
appear to coincide with the result (\ref{finalanswerDus}). In particular,
  terms of the form $x_{12}^2 x_{14}^2 D_{p+2\,p\,5\,5}+x_{23}^2 x_{34}^2 D_{p\,p+2\,5\,5} $
arise from $W_p^5(1,0)$ in (\ref{(5.8)}) and its symmetrization in  $1 \leftrightarrow 3$.
Thanks to the many identities that connect the $D$ functions (see the Appendix),
the two representations of the answer are in fact exactly equal. We first use
 (\ref{asymmetrymajor}) to eliminate the ``asymmetric'' $D$'s in the result
of Section 5.
We get
\bea \label{finalanswerDthem}
I_{\rm grav}& =&
\left( \frac{6}{\pi^2} \right)^4
\left[ 16\, x_{24}^2\left({1\over 2s}-1\right)\,
D_{4455}+{32\over 3} {x_{24}^2\over x_{13}^2}\left(-1+ \frac{2}{3s}\right){x_{24}^2\over x_{13}^2}\, D_{3355} \right.\\ 
  && ~~~~~~~~~~~~~~+
{32 \over 3}{x_{24}^2\over x_{13}^4}\left({1\over2 s}-1\right)\, D_{2255} 
 -\frac{32}{3}\,\frac{x_{24}^2}{x_{13}^6}\,D_{1155} 
+24\,D_{4444} \nonumber \\
&& ~~~~~~~~~~~~~~\left. +
{8\over 9\,x_{13}^2}\,D_{3344}+{14\over 9\,x_{13}^4}\,D_{2244} +{10\over 3\,x_{13}^6}\,D_{1144} \right]
 \,.
\nonumber\eea
Now (\ref{finalanswerDus}), (\ref{finalanswerDthem}) are
both in terms of $D$--functions of the form $D_{\Delta \Delta \tilde \Delta
\tilde \Delta}$. By repeated application of (\ref{scale}) one can convert
one representation into the other. We regard this non--trivial
match as a strong check
of our result.

%%%%%%%%%%%%%%%%%%%%%%%%%%%%%%%%%%%%%%ENDHF%%%%%%%%%%%%%%%%%%%%%%%%%%%%%%%%%%%%%%
\setcounter{equation}{0}

\section{Asymptotic expansions}
We have seen that the graviton exchange amplitude (and generically
all AdS 4--point processes with external scalars) can be expressed
as a finite sum of quartic graphs, see (\ref{finalanswerDus}),
(\ref{(4.18result)}--\ref{(4.19)}), (\ref{finalanswerDthem}).
In this Section we develop 
asymptotic series expansions for the scalar quartic graphs
(Figure 5) 
in terms of conformally invariant variables. This
series expansions allow  to analyze
the supergravity results in terms of the expected double OPE
(\ref{ope}). In Section 3 and 4 we have used slightly different
notations for the quartic graphs, namely $D_{\Delta_1 \Delta_3
\Delta_2 \Delta_4}$ and $W_p^{\Delta}(a,b)$. The connection
between the two is given in (\ref{DW}). Here the expansions
are performed for the $W_p^{\Delta}(a,b)$ representation of the quartic
graph. 

In Section 6.3 we assemble the series expansions
of the $W$'s that appear in the representation (\ref{(3.22)},\ref{(5.8)})
of the graviton exchange for $\Delta = \Delta'=d=4$. We concentrate
on the direct channel and display explicitly  the singular
terms and all the logarithmic contributions. The complete
expansions, in both direct and crossed channels, can be easily
obtained from the formulas in Section 6.2.

\subsection{Integral representations of $W_k^{\Delta'}(a,b)$}

To evaluate $W_k^{\Delta'}(a,b)$, we follow the methods of \cite{dhfgauge} and
 \cite{dhfscalar}. We
introduce a first Feynman parameter $\alpha$ for the denominators $w^2$ and
$(w-x)^2$ and a second Feynman parameter $\beta$ for the resulting denominator
and $(w-y)^2$. The $\vec{w}$ and $w_0$ integrals may then be carried out using
standard formulas, and we find
\bea
W_k^{\Delta '}(a,b)
&=&
{\pi ^\d \over 2} {\Gamma (k+\Delta '+a-\d) \Gamma (\Delta '+b-a) \over
\Gamma (k) \Gamma (\Delta ') \Gamma (\Delta '+b)} \nonumber\\
&& \times 
\int _0 ^1 \! d\alpha \int _0 ^1 \! d\beta
{\alpha ^{\Delta '-1} (1-\alpha)^{k-1} \beta ^{\Delta '+b-1} (1-\beta)^{k+a-b-1}
\over
\bigl [ \beta (y-\alpha x)^2 + \alpha (1-\alpha)x^2\bigr ]
^{\Delta '-a+b}}
 \label{(4.18)}
\eea
Upon performing the following change of variables familiar from 
\cite{dhfgauge} and \cite{dhfscalar}, 
\be
\alpha = {1 \over 1+u} 
\qquad \qquad
\beta = {u \over u+v+uv}
 \label{(4.19bis)}
\ee
we obtain an integral representation similar that of 
 \cite{dhfgauge}
 and \cite{dhfscalar},
\bea
W_k^{\Delta '}(a,b)
&=&
{\pi ^\d \over 2} {\Gamma (k+\Delta '+a-\d) \Gamma (\Delta '+b-a) 
\over
\Gamma (k) \Gamma (\Delta ') \Gamma (\Delta '+b)} 
\int _0 ^\infty \!\! du 
\int _0 ^\infty \!\! dv 
\nonumber\\ &&
\times 
{u^{k+a-1} v^{k+a-b-1} \over (u+v+uv)^{k+2a-b}}
\ 
{1 \over \bigl [ (x-y)^2 + uy^2 +vx^2\bigr
]^{\Delta'-a+b} }
 \label{(4.20)}
\eea

Now the function $W$ with $b\not=0$ only enters the calculation of
$B^{00}$ (equ.(\ref{pippo})), and appears there only in the form of the sum $W_k^{\Delta '}
(1,1)+\tilde W_k^{\Delta'}(1,1)$. This particular combination may be
re--expressed in terms of
$W$-functions with $b=0$ only. This would be difficult to see from the
$w$-integral definition (\ref{(4.13)}), but is manifest from the integral representation
(\ref{(4.20)}), by using the following relation
\be
{u \over (u+v+uv)^{k+1}} + {v \over (u+v+uv)^{k+1}}
= 
{1\over (u+v+uv)^k} - {uv \over (u+v+uv)^{k+1}} 
 \label{(4.21)}
\ee
Taking normalization factors into account properly, we find 
\be
W_k^{\Delta '} (1,1) + \tilde W_k^{\Delta'}(1,1)
=
{k+\Delta '-\d \over \Delta '} W_k ^{\Delta '}(0,0)
-{k\over \Delta '} W^{\Delta '}_{k+1} (0,0)
 \label{(4.22)}
\ee
As a result of this identity, there will be only two classes of $w$-integral
functions entering into the graviton exchange amplitudes : $W_k^{\Delta'}(0,0)$
and $W_k^{\Delta'}(1,0)$.

Similarly, a relation exists expressing $W_k^{\Delta'}(1,0)$ in terms of
$W(0,0)$-functions. This may be established by using the fact that the quantity
\be
\biggl ( {\partial \over \partial u} + {\partial \over \partial v} \biggr )
\biggl ( {(uv)^{k-1} \over (u+v+uv)^k} {1\over \bigl [ (x-y)^2 + uy^2 +vx^2
\bigr ]^{\Delta'}} \biggr )
\ee
has vanishing  integral in $u$ and $v$, and by carrying out the derivatives
explicitly and regrouping the result in terms of $W$-functions. The final
result is
\bea \label{(4.23)}
2(k+1)(\Delta ')^2 W_k^{\Delta '}(1,0)
& =&
k(k+\Delta '-\d)(k+\Delta '-\d-1) W^{\Delta '-1}_k(0,0)
\nonumber\\
&& -k(2k+1)(k+\Delta '-\d) W_{k+1}^{\Delta '-1} (0,0)\\
&& + k(k+1)^2 W_{k+2} ^{\Delta '-1} (0,0)
-k(\Delta ')^2 (x^2+y^2) W_{k+1}^{\Delta'} (0,0)\nonumber
\eea

The $w$-integrals $W_k^{\Delta'}(0,0)$ and $W_k^{\Delta'}(1,0)$ may each be
expressed in terms of derivatives on two universal functions. To show this, we
proceed as in \cite{dhfgauge} and \cite{dhfscalar}, where analogous results were obtained for the scalar
and gauge exchange graphs. We begin by introducing the conformal invariants
\bea
&&s= \half {(x-y)^2 \over x^2 +y^2}=
\frac{1}{2} \frac{x_{13}^2 x_{24}^2}{x_{12}^2 x_{34}^2 + x_{14}^2
x_{23}^2} \\
&&t= {x^2 - y^2 \over x^2 +y^2 }=
 \frac{x_{12}^2 x_{34}^2 -x_{14}^2 x_{23}^2}{x_{12}^2 x_{34}^2
+x_{14}^2 x_{23}^2}
 \label{(4.24)}
\eea
whose ranges are $0\leq s \leq 1$ and $-1\leq t \leq 1$. Next, we perform a
change of variables 
\bea
u & = 2\rho (1-\lambda) \nonumber\\
v & = 2\rho (1+\lambda) 
 \label{(4.25)}
\eea
under which we have 
\bea
W_k^{\Delta '}(a,0)
&=&
{\pi ^\d \over 2^{\Delta '+a}} {\Gamma (k+\Delta '+a-\d) \Gamma (\Delta '-a) 
\over
\Gamma (k) \Gamma (\Delta ')^2 (x^2+y^2)^{\Delta '-a}} 
\int _0 ^\infty \!\! d\rho 
\int _{-1} ^1 \!\! d\lambda 
\nonumber\\ &&
\times 
{\rho ^{k-1} (1-\lambda^2)^{k+a-1} \over [1+\rho (1-\lambda^2)]^{k+2a}}
\ 
{1 \over ( s + \rho +\rho \lambda t ) ^{\Delta'-a} }
 \label{(4.26)}
\eea
It is now possible to write the right hand side as a derivative with respect to
$s$ of order $\Delta '-a-1$ of an integral in which the denominator involving
$s$ appears to degree 1, using
\be
{1 \over (s+\omega)^p} = {(-)^{p+1} \over \Gamma (p)} \biggl (
{\partial \over \partial s} \biggr ) ^{p-1} {1 \over s+\omega}
 \label{(4.27)}
\ee
Next, we change variables to $\rho = s/\mu$ and recognize that the new integral
is a derivative with respect of $s$ of order $k-1 +2a$. Putting all together,
we obtain
\bea
W_k^{\Delta '}(a,0)
&=&
 \pi ^\d  
{(-)^{\Delta '+k+a} 2^{-\Delta '-a} \Gamma (k+\Delta '+a-\d)  
\over
\Gamma (k) \Gamma (k+2a)  \Gamma (\Delta ')^2 (x^2+y^2)^{\Delta '-a}}
\nonumber\\
&& \times
\biggl ( {\partial \over \partial s} \biggr )^{\Delta'-a-1} 
\biggl \{ s^{k-1}
\biggl ( {\partial \over \partial s} \biggr )^{k-1+2a} I_a(s,t) \biggr \}
 \label{(4.28)}
\eea
where the universal functions $I_a(s,t)$ are given by the following integral
representations
\bea
I_a(s,t) & =& s^{2a} \int _0 ^\infty \!\! d\mu 
\int _{-1} ^1 \!\! d\lambda \ 
{(1-\lambda^2)^a \over \mu+ s(1-\lambda^2)} \ {1 \over 1 + \mu + \lambda t}
\nonumber\\
& =&
s^{2a} \int _{-1} ^1 \!\! d \lambda 
{(1-\lambda ^2)^a \over 1+\lambda t -s(1-\lambda ^2)} 
\ln {1+\lambda t \over s(1-\lambda ^2)} 
 \label{(4.29)}
\eea
The integrals $I_a(s,t)$ are perfectly convergent and produce analytic
functions in $s$ and $t$, with logarithmic singularities in $s$ and $t$.

\subsection{Series expansions of  $W_k^{\Delta'}(a,b)$ }

Series expansions of the functions $W^{\Delta'}_k(a,0)$ may be
obtained easily from the series expansions of the universal functions
$I_a(s,t)$. There are two different regions in which the expansion will be
needed : 
\begin{quote}
a) The direct channel (``t--channel'') limit $|x_{13}| \ll |x_{12}|$, $|x_{24}| \ll |x_{12}|$,
which corresponds to $s,t \rightarrow 0$.\\
b) The two crossed channels; one (``s--channel'') is the limit  $|x_{12}| \ll |x_{13}|$,
 $|x_{34}| \ll |x_{13}|$, which corresponds to $s \rightarrow 1/2$,
$t \rightarrow -1$, and the other (``u--channel'') is  $|x_{23}| \ll |x_{34}|$,
  $|x_{14}| \ll |x_{34}|$ in which $s \rightarrow 1/2$, $t \rightarrow 1$.
\end{quote}
We shall now discuss each limit in turn.

\noindent
{\it (a) Direct channel series expansion}

The direct channel limit is given by $s,t \to 0$, and the expansions of the
functions $I_a(s,t)$ are given by 
\bea
I_0 (s,t) & =&
 \sum _{k=0} ^\infty \{ -\ln s \ a_k(t) +b_k(t) \} s^k \label{(4.30a)}\\
I_1 (s,t) & =&
 \sum _{k=0} ^\infty \{ -\ln s \ \hat a_k(t) + \hat b_k(t) \} s^{k+2} \label{(4.30b)}
\eea
where the coefficient functions are given by
\bea
a_k(t) & 
 = {\displaystyle \int _{-1} ^1 \! d\lambda {(1-\lambda ^2) ^k \over (1+\lambda t)^{k+1}}}
\qquad 
{\displaystyle b_k(t) 
 = \int _{-1} ^1 \! d\lambda {(1-\lambda ^2) ^k \over (1+\lambda t)^{k+1}}
  \ln {1+\lambda t \over 1-\lambda ^2}} &\label{(4.31a)}\\
{\displaystyle \hat a_k(t)} & {\displaystyle
 = \int _{-1} ^1 \! d\lambda {(1-\lambda ^2) ^{k+1} \over (1+\lambda t)^{k+1}}}
\qquad
{\displaystyle \hat b_k(t) 
 = \int _{-1} ^1 \! d\lambda {(1-\lambda ^2) ^{k+1} \over (1+\lambda t)^{k+1}}
  \ln {1+\lambda t \over 1-\lambda ^2}} &\label{(4.31b)}
\eea
The coefficient functions admit Taylor series expansions in powers of $t$ with
radius of convergence 1. Actually, in view of (\ref{(4.23)}), we have the following
relations between these functions
\bea
(k+2) \hat a_k (t)\: &
 =& (k+1) \bigl (2a_k(t) - a_{k+1}(t) \bigr )
  \label{(4.32a)} \\
(k+2)^2 \:\hat b_k (t) &
 =& (k+1)(k+2) \bigl (2b_k(t) - b_{k+1}(t) \bigr ) -2a_k(t) + a_{k+1}(t)
  \label{(4.32b)} 
\eea

From (\ref{(4.28)}) and (\ref{(4.30a)}, \ref{(4.30b)}), we obtain the series expansions of $W^{\Delta '}_k(0,0)$
and
$W^{\Delta '} _k (1,0)$ using the following differentiation formulas
\bea
s^p \biggl ( {\partial \over \partial s} \biggr ) ^p  s^k 
&
 = &{\Gamma (k+1) \over \Gamma (k-p+1)} \ s^k 
\label{(4.33a)} \\ 
s^p \biggl ( {\partial \over \partial s} \biggr ) ^p \bigl \{ s^k \ln s \bigr \}
& 
 =& {\Gamma (k+1) \over \Gamma (k-p+1)} \ s^k \ 
\bigl \{ \ln s + \psi (k+1) - \psi (k-p+1) \bigr \} 
 \label{(4.33b)} 
\eea
We find
\bea
W_p^{\Delta }(0,0) 
& =&
  {(-)^{\Delta +p} \pi ^\d \Gamma (p+\Delta -\d) \over 2^\Delta \Gamma (p)^2
  \Gamma (\Delta )^2 (x^2 + y^2 )^\Delta}
\sum _{k=0} ^\infty 
  {\Gamma (k+1)^2 \ s^{k-\Delta +1} \over \Gamma (k-p+2) \Gamma (k-\Delta +2)}
\nonumber\\
&& \biggl \{b_k(t) - a_k(t) \bigl [ \ln s +2 \psi (k+1) - \psi (k-\Delta +2) -\psi (k-p+2)
\bigr ] \biggr \}
 \label{(4.34)}
\eea
and 
\bea \label{(4.35)}
W_p^{\Delta }(1,0) 
& =&
  {(-)^{\Delta +p+1} \pi ^\d \Gamma (p+\Delta -\d+1) 
  \over 2^{\Delta +1} \Gamma (p) \Gamma (p+2)
  \Gamma (\Delta )^2 (x^2 + y^2 )^{\Delta -1}}
\sum _{k=0} ^\infty 
  {\Gamma (k+1) \Gamma (k+3) \ s^{k-\Delta +2} 
  \over \Gamma (k-p+2) \Gamma (k-\Delta +3)}\\
&&\cdot\biggl \{
 \hat b_k(t) - \hat a_k(t) \bigl [ \ln s + \psi (k+1) + \psi (k+3) - \psi
(k-\Delta +3) -\psi (k-p+2)
\bigr ] \biggr \}
 \nonumber
\eea
The presentation of these
 series expansions is slightly formal in the sense that
for $k\leq \Delta -2$, the $\Gamma (k-\Delta +2)$ function in the denominator
produces a zero, while the $\psi (k-\Delta +2)$ term produces a pole, which
together yield a finite result, which amounts to a pole term in $s$. 
Its coefficient can be obtained from the formula
$ \lim_{x \rightarrow 0} \psi(x-q)/ \Gamma(x-q) = (-)^{q+1} \Gamma(q+1)$
for any non--negative interger $q$.

\noindent
{\it (b) Crossed channel series expansion}

The crossed channel asymptotics is given by $s \to \half$ and $t \to \pm 1$,
and may also be obtained from the series expansion of the functions $I_a(s,t)$,
with $a=0,1$. Actually, it suffices to obtain the expansion of $I_0(s,t)$ and
thus of $W^{\Delta'}_p(0,0)$ in this limit and then to compute the series
expansion of $W^{\Delta '}_p(1,0)$ by using the relation (\ref{(4.23)}). This is
useful in this case, since the expansion of $I_1(s,t)$ appears more involved
than that of $I_0(s,t)$.

We start from the definition of $I_0(s,t)$ in (\ref{(4.29)}) as a double integral and
consecutively perform the following changes of variables  $\mu = (1+\lambda
t)\sigma$ and $\tau = (1+\sigma )^{-1}$, so that
\be
I_0(s,t) = \int ^1_0 d\tau \int ^1 _{-1} d \lambda
  {1 \over (1-\tau) (1+\lambda t) + \tau s (1-\lambda^2)}\, .
 \label{(4.36)}
\ee 
This form of the universal function $I_0(s,t)$ is now precisely of the form
studied in \cite{dhfscalar}, and the $\lambda$-integral may be performed explicitly
in an elementary way. We obtain, as in \cite{dhfscalar}
\be
I_0(s,t) = I_0^{{\rm log}}(s,t) + I_0^{{\rm reg}}(s,t)
 \label{(4.37)}
\ee
where
\bea
I_0 ^{{\rm log}} (s,t) & = &-\ln  (1-t^2)  
\int _0 ^1 d \tau {1 \over \sqrt {\omega ^2 -\tau ^2(1-t^2)}}  
 \label{(4.38a)} \\
I_0 ^{{\rm reg}} (s,t) &
 =& 2\int _0 ^1 d \tau {1 \over \sqrt {\omega ^2 -\tau ^2(1-t^2)}}
 \ln \biggl  \{{\omega \over  \tau}  +  \sqrt{{\omega ^2 \over \tau^2}
-(1-t^2)} \biggr \} 
 \label{(4.38b)} 
\eea
where the composite variable $\omega$ is defined by $\omega = 1-(1-2s)(1-\tau)$.
In the neighborhood of $s=\half$ and $t=\pm 1$, we have $\omega \sim 1$ and
$1-t^2 \sim 0$, so that the integrals in (\ref{(4.38a)},\ref{(4.38b)}) 
are both uniformly convergent,
and may be Taylor expanded in powers of $(2s-1)$ and $(1-t^2)$. Thus, $I_0
^{{\rm reg}}(s,t)$ is analytic in both $s$ and $t$ in the neighborhood of
$s=\half$ and $t=\pm 1$, and all non-analyticity is contained in the factor
$\ln (1-t^2)$ of $I^{{\rm log}}_0(s,t)$. The integral admits a double Taylor
expansion given by
\bea
I_0 ^{{\rm log}} (s,t) &
 = &-\ln (1-t^2) \sum _{k=0}^\infty (1-2s)^k \alpha _k(t) \nonumber\\
\alpha _k(t) &
 = &{1 \over k+1} F(\half,{k+1\over 2};{k+3\over 2};1-t^2)
 =\sum _{\ell =0} ^\infty {\Gamma (\ell +\half) \over \Gamma (\half) \ \ell !}
\ {(1-t^2)^\ell \over 2\ell +k+1} 
 \label{(4.39)}
\eea
This expansion may be used to evaluate the logarithmic part of
$W_p^{\Delta'}(0,0)$ and we obtain the following result
\bea
W_p ^{\Delta '}(0,0)\bigg |_{{\rm log}} & =&
 - 2^{p-2} \pi ^\d \ln (1-t^2)
{\Gamma (p +\Delta ' -\d) \over 
 \Gamma (p) \Gamma (\Delta') (x^2 +y^2)^{\Delta '}}\nonumber\\
&& \sum _{\ell =0} ^{\Delta '-1} 
\sum _{k=0}^\infty 
{(-2)^{-\ell} \Gamma(k+1) \ s^{p-\ell-1} \ (1-2s)^{k-p+\ell-\Delta '+2}  
\over 
\Gamma (\Delta '-\ell) \Gamma (p-\ell) \ \ell ! \ \Gamma (k-p+\ell -\Delta '+3)}
\ \alpha _k(t) 
 \label{(4.40)}
\eea
Notice that in the crossed channel, no power singularities arise.

\subsection{Asymptotic expansion for the graviton exchange}

We now turn to the direct channel asymptotic expansion
of the graviton exchange graph for $\Delta = \Delta' = d=4$.
The power singularity terms may be read off directly from the general
asymptotic expansion formula (\ref{(4.34)}) restricted to $d=4$, and we
have
\be
W^\Delta _p (0,0)
\sim 
{ \pi ^2 \Gamma (p+\Delta -2) \over 2^{\Delta }
\Gamma (p)^2 \Gamma (\Delta )^2} {(-)^{p-1} \over 
(x^2 + y^2)^\Delta }
\sum _{k=p-1} ^{\Delta -2} (-)^k
{\Gamma (k+1)^2 \Gamma (\Delta -k-1)  \over \Gamma (k-p+2) } 
{a_k(t) \over s^{\Delta -1-k}}
 \label{(5.9)}
\ee
Similarly, we have from (\ref{(4.35)})
\bea
W^\Delta _p (1,0)
&\sim &
{ \pi ^2 \Gamma (p+\Delta -1) \over 2^{\Delta + 1}
\Gamma (p) \Gamma (p+2) \Gamma (\Delta )^2} {(-)^{p-1} 
\over (x^2 + y^2)^{\Delta -1} }
\nonumber\\
&& \times
\sum _{k=p-1} ^{\Delta -3} (-)^k
{\Gamma (k+1) \Gamma (k+3) \Gamma (\Delta -k-2)  \over \Gamma (k-p+2) }
 {\hat a_k(t) \over s^{\Delta -2-k}}
 \label{(5.10)}
\eea
The full singular power part of the amplitude is now easily obtained by working
out the asymptotics above in the cases $W_p^4 (0,0)$, $W^5_p (0,0)$ and 
$W^5_p(1,0)$ with $p=1,2,3$. The function $W_4^4(0,0)$ has no power
singularities and does not contribute here. Putting all together, we have
\be
B_{{\rm sing}} = -{ 48 \over \pi ^6} {1 \over (x^2+y^2)^4}\biggl [
 -2\bigl (
 {a_0 (t) \over s^3 } + {a_1 (t) \over s^2 } + {a_2 (t) \over s }
 \bigr )
 + 3\bigl (
 {\hat a_0 (t) \over s^3 } + {\hat a_1 (t) \over s^2 } + {\hat a_2 (t) \over s }
 \bigr ) \biggr ]
 \label{(5.11)}
\ee
Using the series expansions of the functions $a_k(t)$ and $\hat a_k(t)$ to low
orders, taking into account that generically, $s$ vanishes like $t^2$,  
\bea
a_0(t) & {\displaystyle = 2 + {2 \over 3} t^2 + {2 \over 5} t^4 
\qquad \quad \
a_1(t) =  {4 \over 3} + {4 \over 5} t^2
\qquad  \quad
a_2 (t) = {16\over 15}}
\nonumber\\
{\displaystyle \hat a_0(t)} & = {\displaystyle {4 \over 3} + {4 \over 15} t^2 + {4 \over 35} t^4 
\qquad
\hat a_1(t) =  {16 \over 15} + {48 \over 105} t^2
\qquad
\hat a_2 (t) = {32\over 35}}
 \label{(5.12)}
\eea 
The final result for the singular part of $B$ is 
\be
B_{\rm sing} = -{2^7 \over  35 \pi ^6 } {1 \over (x^2+y^2)^4} \biggl [
{1 \over s^3} \bigl (-7t^2 -6 t^4 \bigr )
+ {1 \over s^2} \bigl ( 7- 3  t^2 \bigr )
+ {8 \over s} \biggr ]
 \label{(5.13)}
\ee
Repristinating the overall kinematic  factors
we get the final result for the singular terms in
the direct channel of the graviton amplitude
\be
I_{{\rm grav}}\bigg |_{{\rm sing}} =
 {2^{10} \over  35 \pi ^6 } {1 \over x_{13}^8 x_{24}^8} \biggl [
s \,\bigl (7t^2 +6 t^4 \bigr )
+ s^2 \,\bigl (- 7+3  t^2 \bigr )
- 8\,s^3  \biggr]
\ee
Notice that the leading singularity $x_{13}^{-6}$
 cancels between the various tensor
contributions to the amplitude. The physical interpretation of this singular
expansion is discussed in Section 2.3.

The logarithmic singularities may be read off directly from the asymptotic
expansion formulas of (\ref{(4.34)}, \ref{(4.35)}), and we have
\bea
W^4_p(0,0) 
&=&
(-)^{p+1} {\pi ^2 \over 2^6 \ 3^2} {\Gamma (p+2) \over \Gamma (p)^2}
{\ln s \over (x^2+y^2)^4}
\sum _{k=0} ^\infty {\Gamma (k+4) ^2 s^k \over \Gamma (k+5-p) \Gamma (k+1)}
a_{k+3}(t)
\nonumber \\
W^5_p(0,0) 
&=&
(-)^p {\pi ^2 \over 2^{11} \ 3^2} {\Gamma (p+3) \over \Gamma (p)^2}
{\ln s \over (x^2+y^2)^5}
\sum _{k=0} ^\infty {\Gamma (k+5) ^2 s^k \over \Gamma (k+6-p) \Gamma (k+1)}
a_{k+4}(t)
~~~~~~~~~~~~~~~~~\label{(5.14b)} \\
W^5_p(1,0) 
&=&
(-)^{p+1} {\pi ^2 \over 2^{12} \ 3^2} 
{\Gamma (p+4) \over \Gamma (p)\Gamma (p+2)} {\ln s \over (x^2+y^2)^4}
\sum _{k=0} ^\infty {\Gamma (k+6) \Gamma (k+4) s^k \over \Gamma (k+5-p) \Gamma
(k+1)} \hat a_{k+3}(t)  \nonumber 
\eea
Assembling these contributions to the logarithmic singularity and expressing
the coefficient functions $\hat a_k(t)$ in terms of $a_k(t)$ using 
(\ref{(4.32a)}) we get
\bea
I_{\rm grav} \bigg |_{\rm log}  & =& {3\cdot 2^3 \over \pi ^6} {\ln s \over x_{13}^8 x_{24}^8}
\sum _{k=0} ^\infty s^{4+k} \ {\Gamma (k+4) \over \Gamma (k+1)} \bigg \{
-2(5k^2 +20k+16)(3k^2+15k+22) a_{k+3}(t) \nonumber\\&&
 + (k+4)^2 (15k^2 +55 k^2 +42) a_{k+4}(t) \bigg \}\,.
\label{Igravlog}
\eea

\appendix \setcounter{equation}{0}
\section{Appendix \\Properties of $D_{\Delta_1 \Delta_3 \Delta_2 \Delta_4}$}

We have seen that a basic building block in expressing the 4--point functions
is the quantity $D_{\Delta_1 \Delta_3 \Delta_2 \Delta_4}$,
defined by
\be \label{AD}
{D_{\Delta_1 \Delta_3 \Delta_2 \Delta_4}(x_1,x_3,x_2,x_4)=
\int  {d^{d+1}z \over z_0^{d+1}} \,
{\tilde K}_{\Delta_1}(z,x_1)\,{\tilde K}_{\Delta_3}(z,x_3)\, {\tilde K}_{\Delta_2}(z,x_2)\, 
{\tilde K}_{\Delta_4}(z,x_4)}
\ee
where ${\tilde K}_\Delta(z,x)$ is 
\be
\label{AK}{{\tilde K}_\Delta(z,x) = \left( z_0 \over  z_0^2 + (\vec z - \vec x)^2 \right)^\Delta \,.}
\ee 
(note the different normalization from
$K(z,x)$, equ.(\ref{(2.13)})).
Thus $D_{\Delta_1 \Delta_3 \Delta_2 \Delta_4}$  corresponds to a quartic interaction 
between scalars of dimension $\Delta_i$, 
with a simple non--derivative interaction vertex,
see Figure 5. Note that sometimes we suppress the explicit coordinate
dependence of the $D$ functions. Coordinate labels are always
understood to be in the order $(x_1 x_3 x_2 x_4)$. 

While the result of the computation of the graviton exchange graph
gives a sum of many different $D$ functions, in fact all these
functions are closely related to each other. We show 
that one can relate $D_{\Delta_1 \Delta_3 \Delta_2 \Delta_4}$
to $D_{\Delta_1-1 \Delta_3-1 \Delta_2 \Delta_4}$ and $D_{\Delta_1 \Delta_3 \Delta_2+1 
\Delta_4+1}$ (see for example (\ref{scale})). 
Further, all the $D$ functions can be obtained from differentiating
one single expression (which can be obtained in closed form) with 
respect to the variables
$x_{ij}^2$. This is shown in section A.3. Using 
this latter fact we show 
how for example how $D_{\Delta \Delta+1 \tilde \Delta  \tilde \Delta+1}$ (+ symmetrizing
permutations) can be related easily to expressions of the form
$D_{\Delta \Delta \tilde \Delta \tilde \Delta}$ (see (\ref{asymmetry})).
These relations are useful to arrive at
the two simplified forms of the graviton amplitude 
(\ref{finalanswerDus}) , (\ref{finalanswerDthem})
 given in the text and to show their equivalence.
 
\subsection{Relation between $D_{\Delta_1 \Delta_3 \Delta_2 \Delta_4}$ and
$W_k^{\Delta'}(a,b)$}
The standard integral introduced in (\ref{(4.13)}) is just a
quartic graph evaluated in the inverted frame, with some kinematic
factors omitted. The precise relation with  $D_{\Delta_1 \Delta_3 \Delta_2 \Delta_4}$
is
\be \label{DW}
W_k^{\Delta'}(a,b) = x_{13}^{2k}\, x_{14}^{2 \Delta'} \,x_{12}^{2(\Delta'+b)}
 \,D_{2a-b+k,\,  k,\, \Delta'+b , \, \Delta'}
\ee

\subsection{Derivative vertices}
The first thing we note is that if we have 
a quartic interaction with derivatives,
given by a coupling
\be  \phi_{\Delta_1}(z) \,\phi_{\Delta_3}(z)\,{\partial \over {\partial z^\mu}}
\phi_{\Delta_2}(z) \,{\partial \over {\partial z^\nu}}\phi_{\Delta_1}(z)\,g^{\mu \nu}
\, ,
\ee
then the computation of the 4--point function with such an 
interaction can again
be reduced to a sum of terms of the form (\ref{AD}).
 This is done with the identity
\cite{august}
\bea\label{derivident}
g^{\mu \nu} {\partial \over {\partial z^\mu}}{\tilde K}(z,x_1) {\partial \over {\partial z^\nu}}
 {\tilde K}(z,x_2)
&=& \Delta_1 \Delta_2 \left[ 
{\tilde K}_{\Delta_1}(z,x_1) {\tilde K}_{\Delta_2}(z,x_2) \right.
 \\
&& -2 x_{12}^2 {\tilde K}_{\Delta_1+1}(z,x_1) 
\left. {\tilde K}_{\Delta_2+1}(z,x_2) 
\right]\,. \nonumber
\eea
Thus
\bea\label{Dder}
D_{\Delta_1 \Delta_3 \partial \Delta_2 \partial \Delta_4}
& \equiv & \int  {d^{d+1}z \over z_0^{d+1}}{\tilde K}_{\Delta_1}(z,x_1)\,
{\tilde K}_{\Delta_3}(z,x_3)\,  {\partial \over {\partial z^\mu}}{\tilde K}_{\Delta_2}(z,x_2)\,z_0^2\, 
 {\partial \over {\partial z^\mu}}{\tilde K}_{\Delta_4}(z,x_4) \nonumber \\
& =& \Delta_2 \Delta_4 \left(D_{\Delta_1 \Delta_3 \Delta_2 \Delta_4} -2\,x_{24}^2 
\,D_{\Delta_1 \Delta_3 \Delta_2+1 \Delta_4+1}
\right)\,. 
\eea

\subsection{Lowering and raising $\Delta_i$}
Not only does the identity (\ref{Dder}) allow us to remove
derivatives from the quartic vertex, it is also  useful to relate
various $D$ functions to each other. Let us rewrite
the l.h.s. in (\ref{Dder}) as
\bea \label{boxcollection}
D_{\Delta_1 \Delta_3 \partial \Delta_2 \partial \Delta_4} &=&
\frac{1}{2} \int \frac{d^{d+1}z}{z_0^{d+1}} \tilde K_{\Delta_1}(z,x_1)
\tilde K_{\Delta_3}(z,x_3) \Box_z \left( \tilde K_{\Delta_2}(z,x_2) \tilde K_{\Delta_4}(z,x_2)
\right) \nonumber \\
&& -\frac{1}{2} \left(m_{\Delta_2}^2 + m_{\Delta_4}^2   \right)
D_{\Delta_1 \Delta_3  \Delta_2  \Delta_4} 
\eea
where $m_\Delta^2 \equiv \Delta(\Delta-d)$. Upon integrating by parts
of the first term in (\ref{boxcollection}) we get
\bea \label{boxthrowing}
\frac{1}{2} \int [dz] \Box_z \left( \tilde K_{\Delta_1} \tilde K_{\Delta_3} \right)
 \tilde K_{\Delta_2} \tilde K_{\Delta_4} & =& \int [dz] \frac{\partial}{\partial
z^\mu} \tilde K_{\Delta_1} z_0^2 \frac{\partial}{\partial
z^\mu} \tilde K_{\Delta_3} \tilde K_{\Delta_2} \tilde K_{\Delta_4}\nonumber  \\&& +
\frac{1}{2}\left(m_{\Delta_1}^2 + m_{\Delta_3}^2   \right) D_{\Delta_1 \Delta_3  \Delta_2  \Delta_4} 
\eea
Putting relations (\ref{Dder},\ref{boxcollection}, \ref{boxthrowing})
together, we find in particular, for $\Delta_1 = \Delta_3 = \Delta$,
$\Delta_2 = \Delta_4 = \tilde \Delta$:
\be \label{scale}
\tilde \Delta^2 x_{24}^2 D_{\Delta \Delta \tilde \Delta +1 \tilde \Delta+1}
= \Delta^2 x_{13}^2 D_{\Delta+1 \Delta +1\tilde \Delta \tilde \Delta}
+\frac{1}{2}\left( \tilde \Delta^2 -\Delta^2 +m_{\tilde \Delta}^2 - m_{\Delta}
\right) D_{\Delta \Delta \tilde \Delta  \tilde \Delta}
\ee
A special case is $\Delta = \tilde \Delta$, which implies:
\be \label{symmetryminor}
x_{24}^2 D_{\Delta \Delta \Delta+1 \Delta+1} = x_{13}^2
 D_{\Delta+1 \Delta+1 \Delta \Delta}\,.
\ee
Iteration of (\ref{scale}) allows one to prove that  more generally
\be \label{symmetry}
(x_{24}^2)^n \,D_{\Delta \Delta \Delta+n \Delta+n} = (x_{13}^2)^n\,
 D_{\Delta+n \Delta+n \Delta \Delta}\,. 
\ee
\subsection{Obtaining $D_{\Delta_1 \Delta_3 \Delta_2 \Delta_4}$ in closed form}
By using a Schwinger parameterization and performing the $z$ integrals,
one finds \cite{bianchi} (and references therein):
\be \label{Dasintegral}
 D_{\Delta_1 \Delta_3 \Delta_2 \Delta_4}(x_1, x_3, x_2, x_4)=
{{\pi^{d \over 2} \Gamma({{\Sigma -d}\over 2}) \Gamma({\Sigma \over 2})}
\over {2 \,\prod_i \Gamma(\Delta_i)}} \, 
\int  { {\prod_j d\alpha_j \,\alpha_j^{\Delta_j -1}  \delta(\sum_i \alpha_i -1)} \over  
{\left( \sum_{k , l} \alpha_k \alpha_l x_{kl}^2  \right)^{\Sigma \over 2} }}
\ee
where
\be \label{Sigma}
\Sigma \equiv \sum_i \Delta_i \,. 
\ee
We observe that any
 $D_{\Delta_1 \Delta_3 \Delta_2 \Delta_4}$ 
can be obtained by differentiating an appropriate
number of times in the variables $x_{ij}$ the basic function
\be\label{B}
B(x_{ij}) = \int  
{ {\prod_j d\alpha_j \delta(\sum_i \alpha_i -1)} \over {
\left( \sum_{k,l} \alpha_k \alpha_l x_{kl}^2 \right)^2}}\,.
\ee
$B(x_{ij})$ is given in closed form in \cite{bianchi}.
 From the integral representation (\ref{Dasintegral}) we immediately find
\be\label{derivofD}
{\partial \over {\partial x_{13}^2 }} D_{\Delta_1 \Delta_3 \Delta_2 \Delta_4}
= -{ {2 \Delta_1 \Delta_3} \over {\Sigma -d}} \,D_{\Delta_1+1 \Delta_3+1 \Delta_2 \Delta_4}
\ee

\subsection{Symmetrizing identities}

Equation (\ref{derivofD}) can be used to show that 
a sum of $D$ functions which is symmetric under $x_1 \leftrightarrow
x_3$ and $x_2 \leftrightarrow x_4$ can always be rewritten in a basis
in which each individual term shares this symmetry, {\it i.e.}
each term is of the form $D_{\Delta \Delta \tilde \Delta \tilde \Delta}$. 
For example:
\be \label{asymmetry}
x_{12}^2 D_{\Delta+1 \Delta \tilde \Delta +1 \tilde \Delta}
+x_{14}^2  D_{\Delta+1 \Delta \tilde \Delta  \tilde \Delta+1}
= \frac{\Sigma -d}{2 \tilde \Delta} \,
 D_{\Delta \Delta \tilde \Delta  \tilde \Delta}
- \frac{\Delta}{\tilde \Delta} \,x_{13}^2  D_{\Delta+1 \Delta+1 \tilde \Delta \tilde 
\Delta}
\ee 
where $\Sigma \equiv 2 \Delta + 2 \tilde \Delta$. Let us see how to derive
this identity. It follows from conformal invariance
that
\be \label{conformalD}
 D_{\Delta_1 \Delta_3 \Delta_2  \Delta_4} =
\left( \prod_{i < j} (x_{ij}^2)^{-\frac{\Delta_i+\Delta_j}{2} + \frac{\Sigma}{6}}
\right) E_{\Delta_1 \Delta_3 \Delta_2 \Delta_4}(\xi, \eta)
\ee
where
$\xi \equiv \frac{x_{12}^2 x_{34}^2 }{x_{13}^2 x_{24}^2}$, $\eta \equiv 
\frac{x_{14}^2 x_{23}^2}{x_{13}^2 x_{24}^2}$ are conformal cross ratios.
From simple chain rule manipulations we then get
\be \label{chainrule}
\left(
x_{12}^2 \frac{\partial}{\partial x_{12}^2}+x_{13}^2 \frac{\partial}{\partial 
x_{13}^2}+x_{14}^2 \frac{\partial}{\partial x_{14}^2}\right)
 E_{\Delta_1 \Delta_3 \Delta_2 \Delta_4}(\xi, \eta) =0 \,.
\ee
Using (\ref{derivofD}), the last equation is tantamount to (\ref{asymmetry})
for $\Delta_1 =\Delta_3 = \Delta$, $\Delta_2 = \Delta_4 =\tilde \Delta$.
Similar arguments lead to the more complicated identity
\bea \label{asymmetrymajor}
&&x_{12}^2 x_{14}^2 D_{\Delta+2 \,\Delta \tilde \Delta+1 \tilde \Delta+1}    
+x_{23}^2 x_{34}^2 D_{\Delta \Delta+2\, \tilde \Delta+1 \tilde \Delta+1} = \\
&&-\frac{\Delta}{\Delta+1}\,(x_{12}^2x_{34}^2+x_{14}^2x_{23}^2)
 D_{\Delta+1 \Delta+1 \tilde \Delta+1 \tilde \Delta+1} +
\frac{\Delta(\Sigma-d)(\Sigma+2-d)}{4(\Delta+1)\tilde \Delta^2}
 D_{\Delta \Delta \tilde \Delta \tilde \Delta} \nonumber \\
&& -\frac{\Delta(2 \Delta+1)(\Sigma+2-d)}{2(\Delta+1)\tilde \Delta^2}\,
x_{13}^2  D_{\Delta+1 \Delta+1 \tilde \Delta \tilde \Delta} +
\frac{\Delta(\Delta+1)}{\tilde \Delta^2}\, x_{13}^4 
D_{\Delta+2\, \Delta+2\, \tilde \Delta \tilde \Delta} \nonumber \;,  
\eea
where $\Sigma = 2\Delta + 2 \tilde \Delta$.

\subsection{Series expansion of $D_{\Delta \Delta \tilde \Delta
\tilde \Delta}$}
From (\ref{DW}) and (\ref{(4.34)}):
\bea
D_{\Delta \Delta \tilde \Delta
\tilde \Delta} (x_1, x_3, x_2, x_4)
 = 
  {(-)^{\Delta +\tilde \Delta } \pi ^\d \Gamma (\tilde \Delta +\Delta -\d) \over  \Gamma (\tilde \Delta )^2 
\Gamma (\Delta )^2{(x_{13}^2)^\Delta (x_{24}^2)^{\tilde \Delta} } }
\sum _{k=0} ^\infty 
  {\Gamma (k+1)^2 \ s^{k +1} \over \Gamma (k-\tilde \Delta +2) \Gamma (k-\Delta +2)} \nonumber \\
 \biggl \{b_k(t) - a_k(t) \bigl [ \ln s +2 \psi (k+1) - \psi (k-\Delta +2) -\psi (k-\tilde \Delta +2)
\bigr ] \biggr \} \,.~~~~~~~
 \label{Dseries}
\eea

\section*{Acknowledgments}
We are grateful to Edward Witten for important discussions at various
stages of this project.\\
It is a pleasure to acknowledge useful conversations with
Ofer Aharony, Jan de Boer, Richard Brower, Jacques Distler,
Sergio Ferrara, Igor Klebanov,  Juan Maldacena, Joe Polchinski, Alexander Polyakov,
Jacob Sonnenschein,  Leonard Susskind, Alberto Zaffaroni.

The research of E.D'H is supported in part by NSF Grants
No. PHY-95-31023 and PHY-97-22072,
D.Z.F.  by
NSF Grant No. PHY-97-22072, S.D.M., A.M. and L.R. by D.O.E. cooperative agreement
DE-FC02-94ER40818. L.R. is  supported in part by INFN `Bruno Rossi' Fellowship.

\end{document}